\documentclass[a4paper,11pt]{article}
\usepackage{jcappub} 
\pdfoutput=1
\usepackage{enumerate}
\usepackage{amsmath,amssymb}
\usepackage{graphicx}
\usepackage{slashed}
\usepackage{float}
\usepackage[dvipsnames]{xcolor}
\usepackage[normalem]{ulem}
\usepackage{subfigure,orcidlink}
\usepackage{multirow,array}
\usepackage{hyperref}
\hypersetup{%
	colorlinks = true,%
	linkcolor = blue,%
	citecolor = blue,%
	filecolor = blue,%
	urlcolor = blue%
}
\usepackage{tabularray}
\UseTblrLibrary{booktabs}

\title{Cogenesis of visible and dark matter in type-I Dirac seesaw}

\author[a]{Debasish Borah \orcidlink{0000-0001-8375-282X}}
\emailAdd{dborah@iitg.ac.in}
\affiliation[a]{Department of Physics, Indian Institute of Technology Guwahati, Assam 781039, India.}

\author[b]{Partha Kumar Paul \orcidlink{0000-0002-9107-5635}}
\emailAdd{ph22resch11012@iith.ac.in}

\author[b]{and Narendra Sahu \orcidlink{0000-0002-9675-0484}}
\emailAdd{nsahu@phy.iith.ac.in}
\affiliation[b]{Department of Physics, Indian Institute of Technology Hyderabad, Kandi, Telangana-502285, India.}

\abstract{We propose a novel cogenesis framework based on the type-I Dirac seesaw mechanism. The minimal type-I Dirac seesaw with three heavy vector-like fermions $(N)$, one singlet scalar $(\eta)$, and the right-handed counterparts $(\nu_R)$ of the Standard Model (SM) neutrinos is extended to include a Dirac fermion dark matter (DM) $(\chi)$ and its heavier scalar companion ($\phi$). The out-of-equilibrium decays of the vector-like fermion generate asymmetries simultaneously in the visible sector, through decay channels involving $(\nu_R,\eta)$ or lepton, Higgs doublets in the SM, and in the dark sector via decaying into $(\chi,\phi)$. The resulting lepton asymmetry is partially converted into the observed baryon asymmetry by electroweak sphaleron processes, while the dark-sector asymmetry survives to constitute the present-day asymmetric DM relic. The generation of asymmetries in multiple sectors and their mutual washouts provide rich dynamics while also keeping the model testable at different observations involving DM, neutrinos, cosmic microwave background (CMB), as well as gravitational waves (GW). We find that successful cogenesis can be realized for DM masses in the range $100~\mathrm{MeV} \lesssim m_\chi \lesssim 39~\mathrm{TeV}$. The lower bound arises from the requirement that the symmetric component of DM annihilates efficiently before the big bang nucleosynthesis (BBN) epoch, while the upper bound is set by unitarity constraints on the asymmetric DM.}

\keywords{baryon asymmetry, dark matter theory,
	leptogenesis, physics of the early universe}
\arxivnumber{2603.24693}

\begin{document}
	\maketitle
	\flushbottom
	\section{Introduction}\label{sec:intro}
	As suggested by astrophysical and cosmological observations, we live in a Universe whose matter component is dominated by a mysterious, non-luminous, and non-baryonic form of matter known as dark matter (DM) \cite{Planck:2018vyg, ParticleDataGroup:2024cfk}. The sub-dominant or $\sim 20\%$ of the matter density is made up of ordinary or baryonic matter. While the origin of DM is a longstanding puzzle, the asymmetric nature of the baryonic component remains another mystery, referred to as the baryon asymmetry of the Universe (BAU). Although the origins of DM and BAU have been studied separately in several well-motivated frameworks like the weakly interacting massive particle (WIMP) \cite{Kolb:1990vq, Jungman:1995df, Bertone:2004pz} paradigm for DM and baryogenesis \cite{Weinberg:1979bt, Kolb:1979qa}, and leptogenesis \cite{Fukugita:1986hr} for BAU, the similar order of magnitude abundances of DM and BAU, namely, $\Omega_{\rm DM} \approx 5\,\Omega_{\rm B}$ have also led to another pursuit of finding a mechanism of cogenesis that can provide a common origin for both observed phenomena. Some of the widely studied cogenesis mechanisms include asymmetric dark matter (ADM) \cite{Nussinov:1985xr, Davoudiasl:2012uw,Arina:2011cu, Arina:2012fb,Arina:2012aj,Petraki:2013wwa, Zurek:2013wia, Barman:2021ost,Narendra:2017uxl,Narendra:2018vfw,Dutta:2022knf,Mahapatra:2023dbr, Borah:2024wos, Falkowski:2011xh, Falkowski:2017uya, Bandyopadhyay:2025hoc, Borah:2025dka,Asadi:2025vli,Feng:2013zda,Feng:2025wvc,Gu:2009yy,Gu:2009hj,An:2009vq,Blennow:2010qp,Takahashi:2026ngu}, baryogenesis from DM annihilation \cite{Yoshimura:1978ex, Barr:1979wb, Baldes:2014gca, Cui:2011ab, Bernal:2012gv, Bernal:2013bga, Kumar:2013uca, Racker:2014uga, Borah:2018uci, Borah:2019epq, Dasgupta:2019lha, Mahanta:2022gsi, Arora:2025vmw}, Affleck-Dine cogenesis \cite{Roszkowski:2006kw, Seto:2007ym, Cheung:2011if, vonHarling:2012yn, Borah:2022qln, Borah:2023qag}. Some recent works have also proposed cogenesis mechanisms involving a first-order phase transition (FOPT) in the early Universe by utilizing the mass-gain mechanism \cite{Baldes:2021vyz, Azatov:2021irb, Borah:2022cdx, Borah:2023saq}, forbidden decay of DM \cite{Borah:2023god}, or bubble filtering \cite{Borah:2025wzl}.
	
	Motivated by these, here we propose a cogenesis mechanism within the type-I Dirac seesaw framework \cite{Borah:2017dmk}. While type-I seesaw for Majorana neutrinos \cite{Minkowski:1977sc, GellMann:1980vs, Mohapatra:1979ia,Sawada:1979dis,Yanagida:1980xy, Schechter:1980gr} and related cogenesis mechanisms \cite{Falkowski:2011xh} have been studied earlier, the corresponding scenarios with light Dirac neutrinos have not received much attention. Due to the absence of lepton number violation in Dirac neutrino scenarios, our cogenesis mechanism relies on the Dirac leptogenesis paradigm \cite{Dick:1999je, Murayama:2002je, Boz:2004ga,Thomas:2005rs,Gu:2006dc,Bechinger:2009qk,Chen:2011sb,Choi:2012ba,Borah:2016zbd,Gu:2016hxh,Narendra:2017uxl, Borah:2025dka, Bandyopadhyay:2025hoc,Babu:2024glr}. In fact, the idea is very similar to the cogenesis mechanism proposed in \cite{Falkowski:2011xh}. It should be noted that we refer to our model as a ``cogenesis" mechanism purely due to its prediction of equal \textit{CP} asymmetries in dark and lepton sectors. However, this common dynamical origin of equal and opposite \textit{CP} asymmetries does not by itself explain $\Omega_{\rm DM}\approx5\Omega_B$ in all of its realizations. In addition to the respective \textit{CP} asymmetries, the particle-antiparticle asymmetries in the two sectors are governed by independent Yukawa couplings, which control both their production and washout, while the relevant masses in the lepton and dark sectors are unrelated. The mechanism, therefore, provides a shared origin for the visible and dark asymmetries while predicting $\Omega_{\rm DM}/\Omega_{\rm B}$ in a wide range.
	
	In our setup, heavy vector-like fermions taking part in type-I Dirac seesaw are responsible for creating equal and opposite CP asymmetries in the lepton and dark sectors. The presence of three different decay modes of these heavy fermions into right-handed neutrinos, left-handed lepton doublets, and dark fermion leads to rich dynamics compared to the simplest Dirac leptogenesis or Majorana cogenesis models discussed before. Considering hierarchical heavy vector-like fermions, we then solve the coupled Boltzmann equations numerically to calculate the respective asymmetries in the lepton and dark sectors and constrain the parameter space from observational constraints. While the toy model used for our analysis does not provide additional dark sector dynamics, the requirement of annihilating away the symmetric part of DM further constrains the mass of DM. A successful cogenesis also constrains the lightest Dirac neutrino mass, which can have observational implications in experiments sensitive to the absolute neutrino mass scale. 
	
	This paper is organized as follows. In section \ref{sec1}, we briefly discuss our toy model for type-I Dirac cogenesis, followed by the detailed analysis of cogenesis in section \ref{sec2}. We discuss the detection prospects of the model in section \ref{sec3} and finally conclude in section \ref{sec4}.
	
	\section{A minimal setup for Dirac cogenesis}
	\label{sec1}
	We begin by outlining the minimal ingredients required to realize the cogenesis of visible matter and DM within a type-I Dirac seesaw framework. The Standard Model (SM) is extended by three copies of vector-like fermions $N$, one vector-like fermion $\chi$ as DM, three copies of right-handed neutrinos $\nu_R$, and two real scalar fields $\eta$ and $\phi$, all of which are singlets under the SM gauge group. The type-I Dirac seesaw is realized by the presence of heavy fermions $N$, the right chiral part $\nu_R$ of SM neutrinos together with the singlet scalar $\eta$ and the SM Higgs doublet $H$. The DM candidate $\chi$ is accompanied by another dark sector field $\phi$, which facilitates their coupling with the heavy fermion $N$. The out-of-equilibrium decay of $N$ into $LH$ generates a lepton asymmetry in the left-handed sector, while its decay into $\nu_R\eta$ produces a net right-handed asymmetry stored in $\nu_R$. Simultaneously, the decay of $N$ to $\chi\phi$  generates an asymmetric population of $\chi$, which survives as ADM in the present Universe. 
	\begin{figure}[h]
		\centering   \includegraphics[scale=0.23]{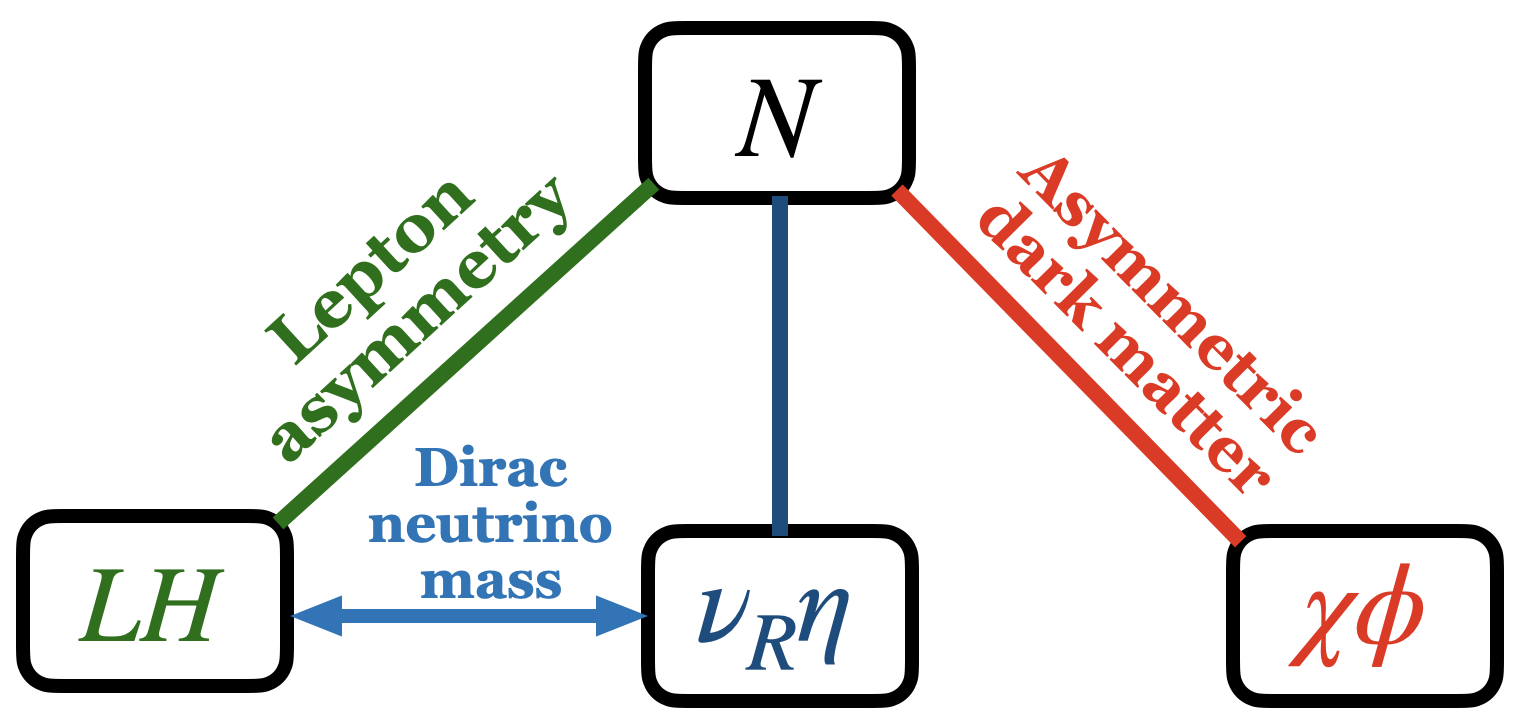}
		\caption{Pictorial representation of our cogenesis scenario in the type-I Dirac seesaw framework.}
		\label{fig:pictorialrep}
	\end{figure}
	To ensure that the present-day DM consists of only an asymmetric component, the symmetric component of $\chi$ must be annihilated away. To ensure that we introduce a light scalar $\phi_1$ and discuss the relevant phenomenology.
	
\begin{figure}[h]
		\centering
		\includegraphics[scale=0.4]{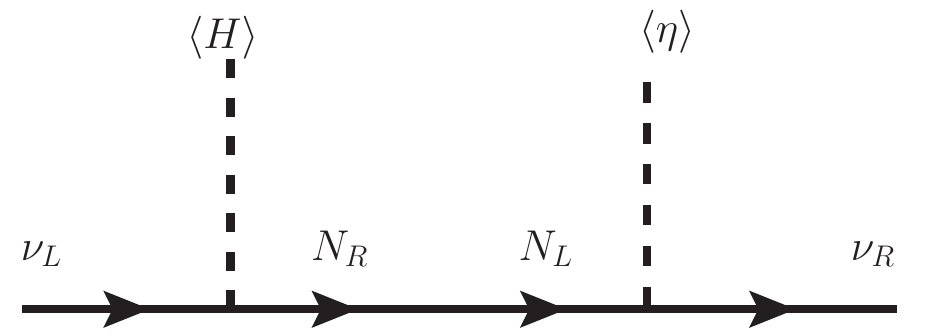}
		\caption{Dirac neutrino mass at tree level.}
		\label{fig:neutrinoFD}
	\end{figure}

	The lepton asymmetry in the left-handed sector is partially converted into the observed baryon asymmetry via electroweak sphaleron processes, thereby explaining the matter–antimatter imbalance of the Universe. The overall setup is illustrated schematically in Fig. \ref{fig:pictorialrep}. Since lepton number is conserved in this framework, light neutrino masses are generated via a Dirac type-I seesaw mechanism involving the fields $L, H, N, \nu_R$ and $\eta$, as depicted in Fig. \ref{fig:neutrinoFD}. The relevant interaction Lagrangian is given as
	\begin{eqnarray}
		\mathcal{L}&=&-y_L\bar{L}\tilde{H}N_i- y_R\bar{N_i}\eta\nu_R-y_{\chi_i} \bar{N_i}\chi\phi -y_{\phi_1}\bar{\chi}\chi\phi_1+{\rm h.c.}
	\end{eqnarray}
	Such interactions arise naturally upon imposing some suitable symmetry $\mathcal{G}$ which also stabilizes DM and forbids other unwanted couplings like $\bar{L} \tilde{H} \nu_R$. We present an explicit model realizing this framework in Appendix \ref{app:model}. 
	
	The symmetry that restricts the interaction $LH\nu_R$ is either softly or spontaneously broken by $\eta$ when it acquires a non-zero vacuum expectation value (VEV). 
	After the electroweak symmetry breaking, the light Dirac neutrino mass can then be estimated from Fig. \ref{fig:neutrinoFD} as
	\begin{eqnarray}
		m_\nu\simeq \frac{y_Ly_Rv_hv_\eta}{2m_{N}},
	\end{eqnarray}
	where $m_N$ is the mass of $N$ and $v_h$ is the SM Higgs VEV. If $\eta$ acquires an induced VEV due to a soft symmetry breaking term like $\mu_1 \eta H^\dagger H$, then one can write its VEV as $v_\eta\simeq\mu_1v_h^2/2m_\eta^2$ where $m_\eta$ is the mass of $\eta$. The details of the parameterization of the left and right-handed sector Yukawa couplings $y_L$ and $y_R$ are given in Appendix \ref{app:yukawapara}. Diagonalizing the above mass matrix, we get the physical light neutrino masses as $m_1,m_2$ and $m_3$ as discussed in Appendix \ref{app:yukawapara}.

	\section{Cogenesis of visible and dark matter}
	\label{sec2}
\begin{figure}[h]
		\centering
		\includegraphics[scale=0.8]{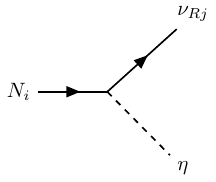}
		\includegraphics[scale=0.8]{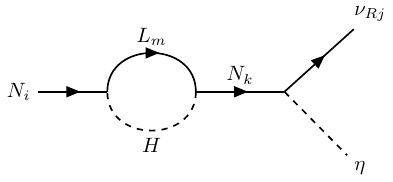}
		\includegraphics[scale=0.8]{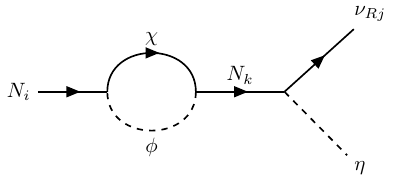}\\
		\includegraphics[scale=0.8]{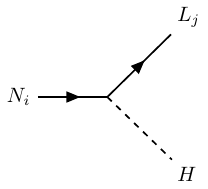}
		\includegraphics[scale=0.8]{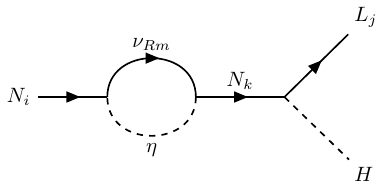}
		\includegraphics[scale=0.8]{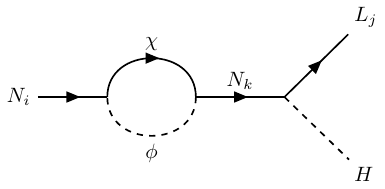}\\
		\includegraphics[scale=0.8]{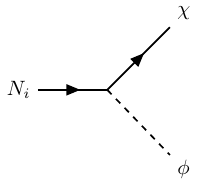}
		\includegraphics[scale=0.8]{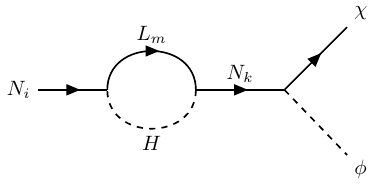}
		\includegraphics[scale=0.8]{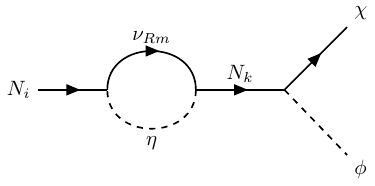}
		\caption{Decay of $N$ contributing to the non-zero \textit{CP} asymmetries: $\epsilon_R$ (top), $\epsilon_L$ (middle), and $\epsilon_\chi$ (bottom).}
		\label{fig:cpR}
	\end{figure}

	The out-of-equilibrium decays of the vector-like fermion $N$ into $LH$, $\nu_R\eta$, and $\chi\phi$ simultaneously generate asymmetries in the $L$, $\nu_R$, and $\chi$ sectors. The interference between the tree-level and one-loop decay amplitudes, as shown in Fig. \ref{fig:cpR}, gives rise to a non-zero CP asymmetry in each channel, namely $\epsilon_L,\epsilon_R$ and $\epsilon_\chi$, respectively. However, since total lepton number is conserved due to the purely Dirac nature of the fermions, the total CP asymmetry summed over all decay modes from $N_i$ decay vanishes, i.e.,
	\begin{eqnarray}
		\epsilon^i_L+\epsilon^i_R+\epsilon^i_\chi=0.
	\end{eqnarray}
	As no lepton number-violating vertex is available, no vertex correction to the $N_i$ decays can be constructed. The \textit{CP} asymmetries thus arise solely from the $N_i$ self-energy diagrams of Fig. \ref{fig:cpR}.
	The corresponding \textit{CP} asymmetries, assuming hierarchical $N$'s, are given as
	\begin{eqnarray}
		\epsilon^i_R\simeq\frac{1}{8\pi}\sum_{k}\frac{m_{N_i}}{m_{N_k}}\frac{{\rm Im}\left[ (y_R^\dagger y_R)_{ki} \left\{ (y_L^\dagger y_L)_{ik} + y^*_{\chi i} y_{\chi k} \right\}  \right]}{(y_R^\dagger y_R)_{ii}+(y_L^\dagger y_L)_{ii}+y^*_{\chi i} y_{\chi i}},
	\end{eqnarray}
	\begin{eqnarray}
		\epsilon^i_L\simeq\frac{1}{8\pi}\sum_{k}\frac{m_{N_i}}{m_{N_k}}\frac{{\rm Im}\left[ (y_L^\dagger y_L)_{ki} \left\{ (y_R^\dagger y_R)_{ik} + y^*_{\chi i} y_{\chi k} \right\}  \right]}{(y_R^\dagger y_R)_{ii}+(y_L^\dagger y_L)_{ii}+y^*_{\chi i} y_{\chi i}},
	\end{eqnarray}
	\begin{eqnarray}
		\epsilon^i_\chi\simeq\frac{1}{8\pi}\sum_{k}\frac{m_{N_i}}{m_{N_k}}\frac{{\rm Im}\left[  \left\{(y_R^\dagger y_R)_{ik} +(y_L^\dagger y_L)_{ik}\right\} y_{\chi i} y^*_{\chi k}  \right]}{(y_R^\dagger y_R)_{ii}+(y_L^\dagger y_L)_{ii}+y^*_{\chi i} y_{\chi i}}.
	\end{eqnarray}
	
	To track the evolution of asymmetries in the left-handed sector, right-handed sector, DM, as well as $N_1$ abundance, we solve the following sets of coupled Boltzmann equations
	\begin{eqnarray}
		\frac{dY_{N_1}}{dz}&=&-\frac{1}{n_\gamma \mathcal{H}z} \left( \frac{Y_{N_1}}{Y^{\rm eq}_{N_1}}-1 \right)\left[ \gamma(N_1\rightarrow LH)+\gamma(N_1\rightarrow \nu_R\eta)+\gamma(N_1\rightarrow \chi\phi) \right]\label{eq:YN1},
	\end{eqnarray}
	\begin{eqnarray}
		\frac{dY_{\Delta\nu_R}}{dz}&=&\frac{1}{n_\gamma \mathcal{H}z}\left[\epsilon_R \left( \frac{Y_{N_1}}{Y^{\rm eq}_{N_1}}-1 \right) \gamma(N_1\rightarrow \nu_R\eta)-\frac{1}{2}\frac{Y_{\Delta\nu_R}}{Y^{\rm eq}_{\nu_R}}\gamma(N_1\rightarrow \nu_R\eta)+ \right.\nonumber\\&&\left. \left( \frac{Y_{\Delta\chi}}{Y^{\rm eq}_{\chi}}-\frac{Y_{\Delta\nu_R}}{Y^{\rm eq}_{\nu_R}}\right) \gamma(\chi\phi\rightarrow \nu_R\eta) - \left(\frac{Y_{\Delta\nu_R}}{Y^{\rm eq}_{\nu_R}}-\frac{Y_{\Delta L}}{Y^{\rm eq}_{L}}\right) \gamma(\nu_R\eta\rightarrow LH)  \right]\label{eq:YnuR},
	\end{eqnarray}
	\begin{eqnarray}
		\frac{dY_{\Delta L}}{dz}&=&\frac{1}{n_\gamma \mathcal{H}z}\left[\epsilon_L \left( \frac{Y_{N_1}}{Y^{\rm eq}_{N_1}}-1 \right) \gamma(N_1\rightarrow LH)-\frac{1}{2}\frac{Y_{\Delta L}}{Y^{\rm eq}_{L}}\gamma(N_1\rightarrow LH)+ \right.\nonumber\\&&\left. \left( \frac{Y_{\Delta\nu_R}}{Y^{\rm eq}_{\nu_R}}-\frac{Y_{\Delta L}}{Y^{\rm eq}_{L}}\right) \gamma(\nu_R\eta\rightarrow LH) +\left( \frac{Y_{\Delta\chi}}{Y^{\rm eq}_{\chi}}-\frac{Y_{\Delta L}}{Y^{\rm eq}_{L}}\right) \gamma(\chi\phi\rightarrow LH)  \right]\label{eq:YL},
	\end{eqnarray}
	\begin{eqnarray}
		\frac{dY_{\Delta \chi}}{dz}&=&\frac{1}{n_\gamma \mathcal{H}z}\left[\epsilon_\chi \left( \frac{Y_{N_1}}{Y^{\rm eq}_{N_1}}-1 \right) \gamma(N_1\rightarrow \chi\phi)-\frac{1}{2}\frac{Y_{\Delta \chi}}{Y^{\rm eq}_{\chi}}\gamma(N_1\rightarrow \chi\phi) \right.\nonumber\\&&\left. -\left(\frac{Y_{\Delta \chi}}{Y^{\rm eq}_{\chi}}-\frac{Y_{\Delta\nu_R}}{Y^{\rm eq}_{\nu_R}}\right) \gamma(\chi\phi\rightarrow \nu_R\eta) - \left(\frac{Y_{\Delta\chi}}{Y^{\rm eq}_{\chi}}- \frac{Y_{\Delta L}}{Y^{\rm eq}_{L}}\right) \gamma(\chi\phi\rightarrow LH)  \right]\label{eq:Yx},
	\end{eqnarray}
	
	where $z=m_{N_1}/T$, $Y_x=n_x/n_\gamma$, with $n_\gamma$ being the photon number density. In the above equations, $\gamma(N_1\rightarrow ab)$ denotes the thermally averaged reaction density for the decay of $N_1$ to various modes. We denote $\epsilon_R\equiv\epsilon_R^1$, $\epsilon_L\equiv\epsilon_L^1$, and $\epsilon_\chi\equiv\epsilon_\chi^1$. $\gamma(ab\rightarrow cd)$ denotes the reaction density for processes that transfer asymmetries between different sectors, thereby acting as washout processes for the corresponding asymmetries. The details of the cross-sections and the decay widths are provided in Appendix \ref{app:crossec}.
	
	As the asymmetry is generated in the left-handed sector, it is partially converted into the baryon asymmetry through electroweak sphaleron processes and can be expressed as
	\begin{eqnarray}   \eta_B(z\rightarrow\infty)=\frac{C_{\rm sph}}{f}Y_{\Delta L}(z\rightarrow\infty),
	\end{eqnarray}
	$C_{\rm sph}=28/79$ is the sphaleron conversion factor\footnote{In our setup, $\nu_R$ is in equilibrium through the interaction among the dark sector particles such as $\nu_R\eta\leftrightarrow \nu_R\eta$ and $\nu_R\eta\leftrightarrow \chi\phi$ mediated by $N_i$. Being an $SU(2)_L$ singlet with zero hypercharge, and with the $\bar{L}\tilde{H}\nu_R$ operator forbidden by symmetry, $\nu_R$ does not enter the chemical equilibrium conditions of the sphaleron processes. The standard conversion factor 28/79, therefore, remains unmodified\cite{Dick:1999je,Cerdeno:2006ha,Heeck:2023soj,Ahmed:2025vzl,Babu:2024glr}.} \cite{Harvey:1990qw}, $f\simeq115.75/3.91\simeq29.6$ is the entropy dilution factor.
	
	It is important to note that the DM asymmetry $Y_{\Delta\chi}$ is not a directly observable quantity. The relevant observable is the DM relic abundance $\Omega_{\chi}h^2=0.12 \pm 0.0012$. We can relate the asymmetries in the visible and dark sectors to this observed quantity as
	\begin{eqnarray}
		\mathcal{R}\equiv\frac{\Omega_\chi h^2}{\Omega_B h^2}=\frac{f}{C_{\rm sph}}\frac{m_\chi}{m_p}\frac{Y_{\Delta\chi}}{Y_{\Delta L}}=5.364 \pm 0.065,
	\end{eqnarray}
	where $\Omega_Bh^2=0.02237 \pm 0.00015$ is the baryon relic density, and $m_p$ is the proton mass. Assuming the symmetric part of DM to annihilate away, the net DM abundance can be parameterized as
	\begin{eqnarray}
		\Omega_{\chi}h^2\sim0.12\left( \frac{m_\chi}{1\,{\rm GeV}} \right)\left( \frac{Y_{\Delta\chi}}{2.9995\times10^{-9}} \right).
	\end{eqnarray}
	
	\subsection{Unitarity bound on the asymmetric dark matter mass}
	Thermal DM mass is bounded from above by the model-independent unitarity limit of $\mathcal{O}(100 \, \rm TeV)$ \cite{Griest:1989wd}. Similar but stricter upper bounds exist on asymmetric DM mass, too, from the requirement of the symmetric part to have a negligible relic abundance \cite{Baldes:2017gzw}. We briefly summarize this bound on ADM mass below.
	
	The DM mass can be written in terms of the relic abundance of DM and relic abundance of baryons at the present epoch as \cite{Baldes:2017gzw}
	\begin{eqnarray}
		m_{\rm DM}\equiv m_{\chi}=\frac{m_p}{\kappa}\frac{\Omega_{\chi}}{\Omega_B}\left( \frac{1-r_\infty}{1+r_\infty} \right),
	\end{eqnarray}
	where $r_\infty=Y_\chi^-/Y_\chi^+$ is the ratio of the anti-DM abundance to the DM abundance, and $\kappa=Y_{\Delta\chi}/\eta_B$. Then $r_{\infty}$ can be expressed as 
	\begin{eqnarray} r_\infty=\frac{\frac{m_p\mathcal{R}}{\kappa}-m_{\rm DM}}{\frac{m_p\mathcal{R}}{\kappa}+m_{\rm DM}}.
	\end{eqnarray}
	For a non-zero $\kappa$ and in the limit $r_\infty\rightarrow0$, we have a maximum limit on the DM as
	\begin{eqnarray}
		m_{\rm DM}<m_{\rm max}=\frac{m_p}{\kappa}\frac{\Omega_{\rm DM}}{\Omega_B}=\frac{m_p\mathcal{R}}{\kappa}\simeq\frac{5{~\rm GeV}}{\kappa}.
	\end{eqnarray}
	To realize this, we need an infinitely large annihilation cross-section which can ensure $r_\infty\rightarrow0$. Alternatively, partial-wave unitarity places an upper bound on the inelastic cross-section, thereby implying $r_\infty\neq0$ but very small. Consequently, this leads to a stronger upper bound on the mass of ADM as
	\begin{eqnarray}
		m_{\rm DM}<m_{\rm uni}<m_{\rm max}.
	\end{eqnarray}
	The unitarity limit on the ADM mass is given as \cite{Baldes:2017gzw}
	\begin{eqnarray}
		m_{\rm uni}\approx m^{\rm sym}_{\rm uni}\left[ \left( \frac{1+r_\infty}{1-r_\infty} \right) \frac{\ln(1/r_\infty)}{2} \right]^{-1/2},
	\end{eqnarray}
	with
	\begin{equation}
		m_{\rm uni}^{\rm sym}\approx\begin{cases}
			110{~\rm TeV}, & \text{for $s$-wave}.\\
			110\sqrt{3}{~\rm TeV}, & \text{for $p$-wave}.
		\end{cases}
	\end{equation}
	For very small $r_\infty$ the above expression can be estimated to be
	\begin{eqnarray}
		m_{\rm uni}\simeq m^{\rm sym}_{\rm uni}\sqrt{2}\left[\ln(1/r_\infty)\right]^{-1/2}
	\end{eqnarray}
	Then the unitarity for the ADM can be approximated as
	where
	\begin{equation}
		m_{\rm uni}(r_\infty)\approx\begin{cases}
			110\sqrt{2}\left[\ln(1/r_\infty)\right]^{-1/2}{~\rm TeV}, & \text{for $s$-wave}.\\
			110\sqrt{6}\left[\ln(1/r_\infty)\right]^{-1/2}{~\rm TeV}, & \text{for $p$-wave}.
		\end{cases}
	\end{equation}

	\subsection{Numerical solution of the Boltzmann equation}
	
	The model contains several free parameters associated with the neutrino sector and the dark sector. To explore the viable region of the parameter space, we perform a Markov Chain Monte Carlo (MCMC) scan over the model parameters. See Appendix \ref{app:mcmc} for the details on MCMC setup. In order to reduce the dimensionality of the scan, we assume $y_{\chi_1}=y_{\chi_2}=y_{\chi_3}\equiv y_\chi$, the dark singlet scalar ($\phi$) mass $m_\phi=5\times m_\chi$
	and fix the singlet scalar ($\eta$) mass to $m_\eta=1~{\rm GeV}$\footnote{We note that as long as the scalar masses are much smaller than $N_1$, the results remain unchanged. We also note that the choice of a general dark sector couplings: $y_{\chi_1} \neq y_{\chi_2} \neq y_{\chi_3}$, does not alter the overall conclusion of the study.}. 
	With these simplifying assumptions, the effective dimensionality of the parameter space is reduced to 9, namely \{$m_1,m_{N_1},m_{N_2}/m_{N_1},m_{N_3}/m_{N_2},m_\chi,v_\eta,a,b,y_\chi$\}. The Yukawa coupling matrices entering the neutrino mass generation are parameterized using the Casas-Ibarra (CI) parametrization, as described in Appendix~\ref{app:yukawapara}. This parametrization allows us to reconstruct the Yukawa matrices $y_L$ and $y_R$ in terms of the low-energy neutrino parameters and a complex orthogonal matrix $R$, which we parameterize using a complex angle $\theta = a + i b$. In the numerical scan, we fix the neutrino mass-squared differences and mixing angles to their experimentally measured best-fit values \cite{deSalas:2020pgw}, while the lightest neutrino mass $m_1$ is treated as a free parameter. We solve the set of coupled Boltzmann equations numerically to determine the final baryon asymmetry and DM relic abundance. Before delving into the full parameter scan, we first consider four representative benchmark points, with DM masses spanning the MeV to multi-TeV range. These points are listed in Table \ref{tab:tab0}, and the corresponding evolution of the asymmetries is presented for illustration.
	
	\begin{figure}[h]
		\centering
		\includegraphics[scale=0.35]{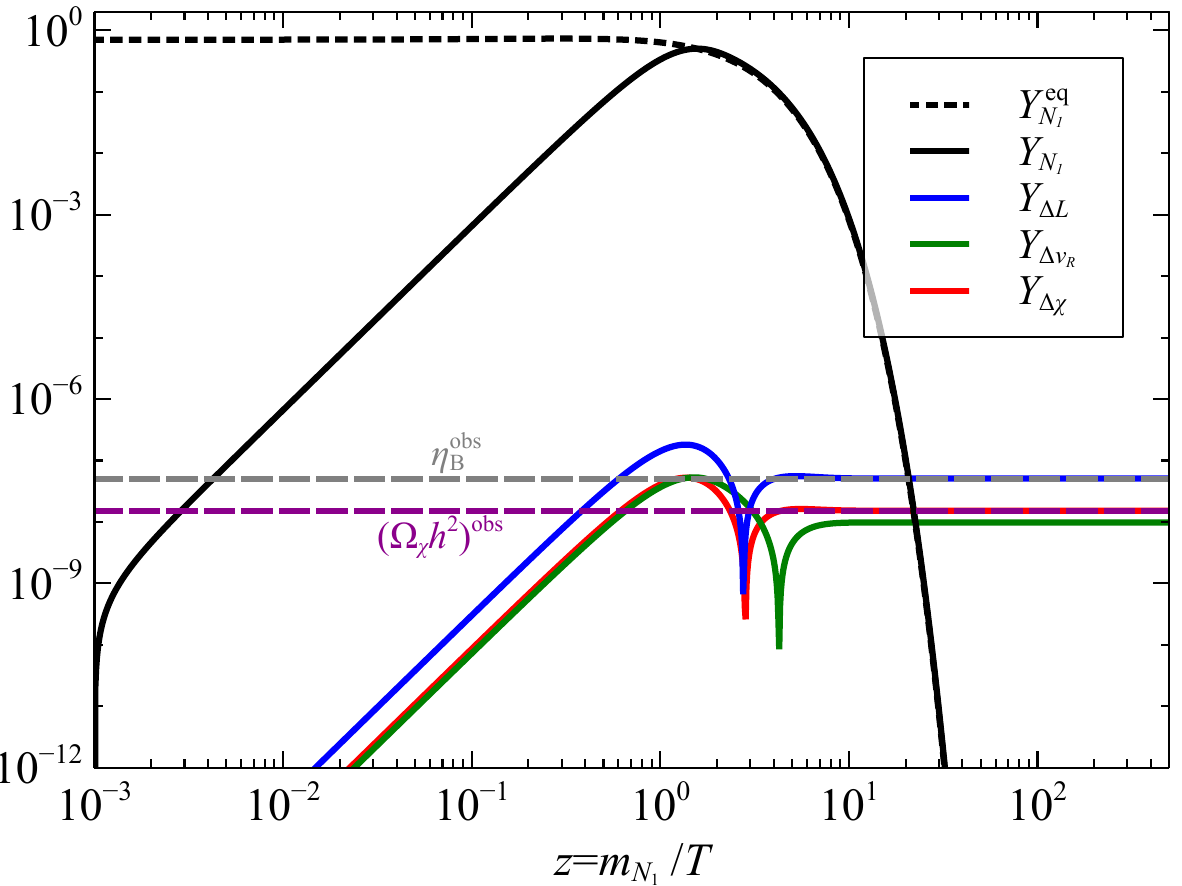}
		\includegraphics[scale=0.35]{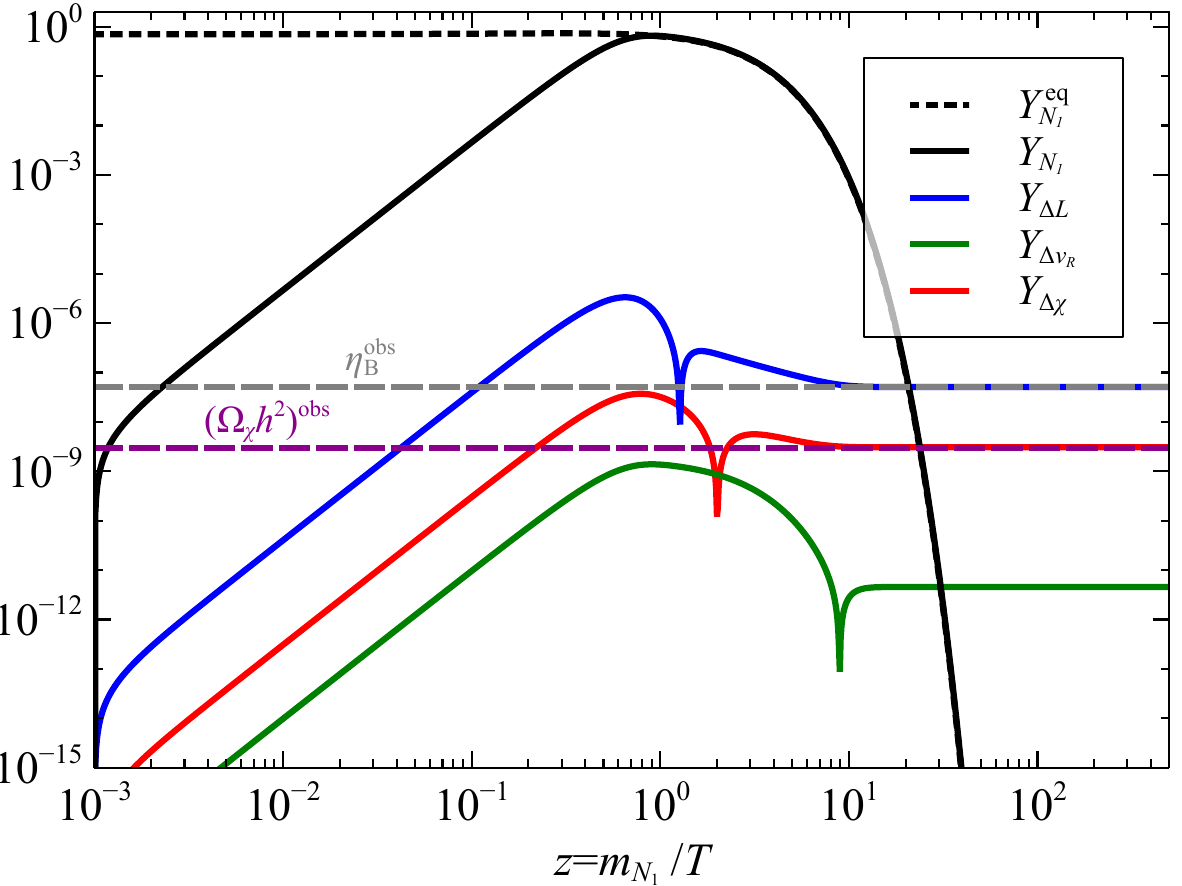}
		\caption{\textit{[Left]}: Cosmological evolutions of the $N_1$ abundance, along with the left-handed, right-handed visible sector asymmetries and dark sector asymmetry, are shown for the BP1. The \textit{CP} asymmetry parameters are given as $\epsilon_{R}=-1.252\times10^{-6}$, $\epsilon_{L}=9.395\times10^{-7}$, $\epsilon_{\chi}= 3.127\times10^{-7}$. The final baryon asymmetry is obtained to be $\eta_B=6.08\times10^{-10}$ and the ratio of the DM relic to baryon relic is $\mathcal{R}=5.365$. The left-handed, right-handed, and DM asymmetries are given as $Y_{\Delta{L}}=5.076885\times10^{-8}$, $Y_{\Delta{\nu_R}}=9.808319\times10^{-9}$, $Y_{\Delta{\chi}}=1.520480\times10^{-8}$, respectively. \textit{[Right]}: The same for the BP2. The \textit{CP} asymmetry parameters are given as $\epsilon_{R}=1.061\times10^{-5}$, $\epsilon_{L}=-1.017\times10^{-5}$, $\epsilon_{\chi}= -4.384\times10^{-7}$.  The final baryon asymmetry is obtained to be $\eta_B=6.094678\times10^{-10}$ and the ratio of the DM relic to baryon relic is $\mathcal{R}=5.35$. The left-handed, right-handed, and DM asymmetries are given as $Y_{\Delta{L}}=5.089927\times10^{-8}$, $Y_{\Delta{\nu_R}}=4.549656\times10^{-12}$, $Y_{\Delta{\chi}}=3.056264\times10^{-9}$, respectively.}
		\label{fig:bp12}
	\end{figure}
	
	In the \textit{left} panel of Fig. \ref{fig:bp12}, we show the cosmological evolution of the asymmetries and $N_1$ abundance for BP1, as given in Table \ref{tab:tab0}. This BP corresponds to $m_{\chi}=201.2184$ MeV. For BP1, the \textit{CP} asymmetries are found to be $\epsilon_{R}=-1.252\times10^{-6}$, $\epsilon_{L}=9.395\times10^{-7}$, and $\epsilon_{\chi}= 3.127\times10^{-7}$, which satisfy the $\epsilon_L+\epsilon_R+\epsilon_\chi=0$ condition. The $N_1$ abundance is shown with the black solid line, whereas the black dashed line represents its equilibrium value. The blue solid line depicts the lepton asymmetry. The green solid line corresponds to the asymmetry generated in $\nu_R$, and the DM asymmetry is shown with the red solid line. The observed values of the lepton asymmetry and DM abundance are shown with gray and magenta dashed lines, respectively. The ADM abundance is found to be $Y_{\Delta\chi}=1.52\times10^{-8}$.
	
	\begin{figure}[h]
		\centering
		\includegraphics[scale=0.35]{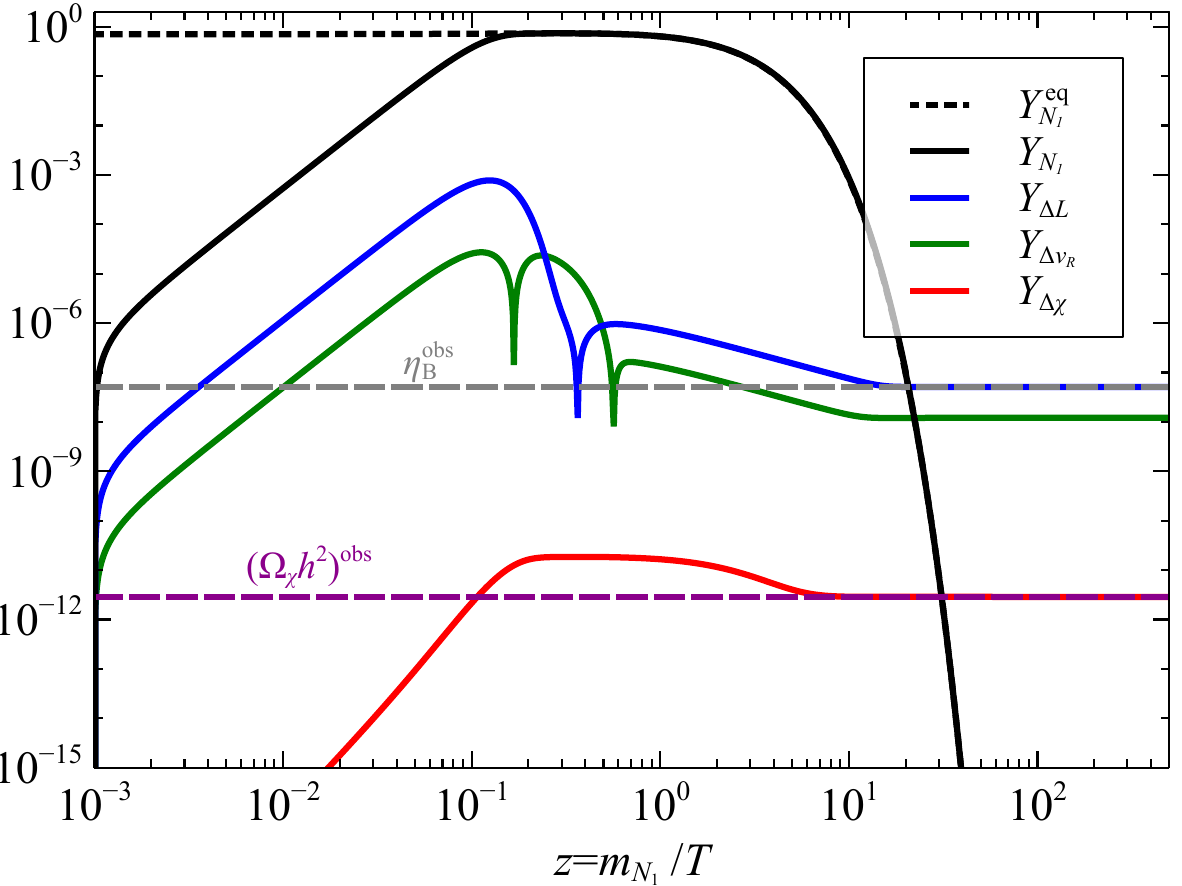}
		\includegraphics[scale=0.35]{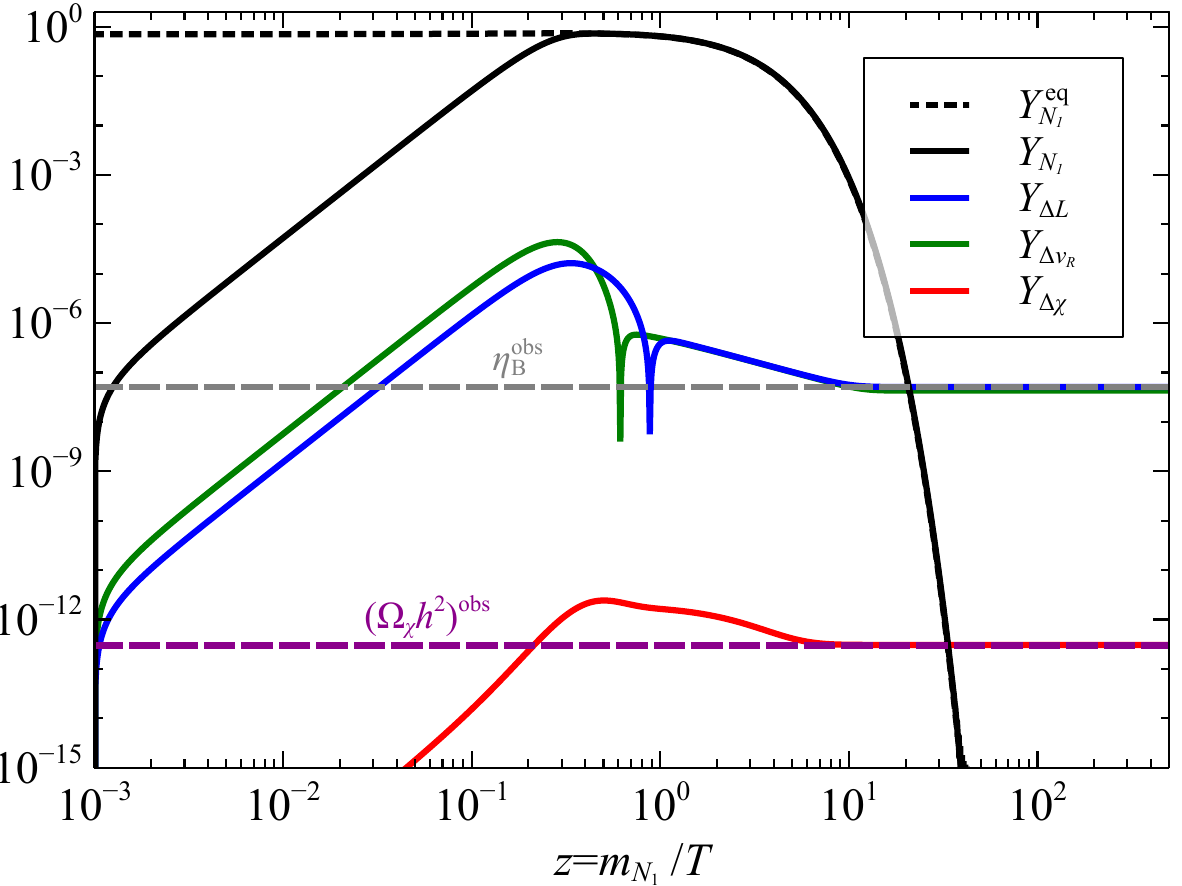}
		\caption{\textit{[Left]}: Cosmological evolutions of the $N_1$ abundance, along with the left-handed, right-handed visible sector asymmetries and dark sector asymmetry, are shown for the BP3. The \textit{CP} asymmetry parameters are given as $\epsilon_{R}=-2.221\times10^{-3}$, $\epsilon_{L}=2.221\times10^{-3}$, $\epsilon_{\chi}= -1.089\times10^{-9}$. The final baryon asymmetry is obtained to be $\eta_B=6.09338\times10^{-10}$ and the ratio of the DM relic to baryon relic is $\mathcal{R}=5.35$. The left-handed, right-handed, and DM asymmetries are given as $Y_{\Delta{L}}=5.088843\times10^{-8}$, $Y_{\Delta{\nu_R}}=1.209899\times10^{-8}$, $Y_{\Delta{\chi}}=2.871683\times10^{-12}$, respectively. \textit{[Right]}: The same for the BP4. The \textit{CP} asymmetry parameters are given as $\epsilon_{R}=-1.371\times10^{-4}$, $\epsilon_{L}=1.371\times10^{-4}$, $\epsilon_{\chi}= -1.578\times10^{-9}$. The final baryon asymmetry is obtained to be $\eta_B=6.094436\times10^{-10}$ and the ratio of the DM relic to baryon relic is $\mathcal{R}=5.4$. The left-handed, right-handed, and DM asymmetries are given as $Y_{\Delta{L}}=5.089725\times10^{-8}$, $Y_{\Delta{\nu_R}}=4.307413\times10^{-8}$, $Y_{\Delta{\chi}}=3.056632\times10^{-13}$, respectively.}
		\label{fig:bp34}
	\end{figure}
	
	We now turn to BP2, for which the DM mass is $m_\chi = 1~\mathrm{GeV}$. The corresponding $CP$ asymmetry parameters are found to be $\epsilon_R = 1.061\times10^{-5}$, $\epsilon_L = -1.017\times10^{-5}$, and $\epsilon_\chi = -4.384\times10^{-7}$. Since the DM mass is larger than that of the previous benchmark point, a smaller ADM abundance is sufficient to reproduce the observed relic density. For this benchmark point, we obtain a final comoving DM asymmetry of $Y_{\Delta\chi} = 3.056\times10^{-9}$.
	
	In the \textit{left} panel of Fig.~\ref{fig:bp34}, we show the evolution of the asymmetries for BP3, where the DM mass is $m_\chi = 1.06~\mathrm{TeV}$. The corresponding $CP$ asymmetry parameters are $\epsilon_R = -2.221\times10^{-3}$, $\epsilon_L = 2.221\times10^{-3}$, and $\epsilon_\chi = -1.089\times10^{-9}$. The final ADM yield is found to be $Y_{\Delta\chi} = 2.872\times10^{-12}$.
	
	Finally, the evolution of the asymmetries for BP4 is shown in the \textit{right} panel of Fig.~\ref{fig:bp34}. In this case, the DM mass is $m_\chi = 10.1~\mathrm{TeV}$. The corresponding $CP$ asymmetry parameters are $\epsilon_R = -1.371\times10^{-4}$, $\epsilon_L = 1.371\times10^{-4}$, and $\epsilon_\chi = -1.578\times10^{-9}$. The resulting final ADM yield is $Y_{\Delta\chi} = 3.057\times10^{-13}$.
	
	\begin{table}[h]
		\centering
		\resizebox{\textwidth}{!}{
			\begin{tabular}{|c|c|c| c|c| c|c|c|c|c|c|} 
				\hline
				BPs &$m_{1}$ (eV) & $m_{N_1}$ (GeV) & $m_{N_2}$ (GeV) & $m_{N_3}$ (GeV) & $m_{\chi}$ & $\theta=a+ib$ & $v_\eta$ (GeV) & $y_{\chi_{1,2,3}}$ \\
				\hline
				BP1   & $3.866295\times10^{-10}$ &$4.05271\times10^{8}$ & $5.804876\times10^{8}$ & $1.456129\times10^{11}$ & 201.2184 MeV & $-0.5285583+i0.3594082$ &$1.208055$ & $2.544335\times10^{-4}$ \\
				\hline
				BP2   & $1.216875\times10^{-10}$ &$1.518230\times10^{10}$ & $2.935591\times10^{10}$ & $5.300304\times10^{12}$ &$1$ GeV & $5.159827-i1.286167$ & 22.00991 & $2.474511\times10^{-3}$ \\
				\hline
				BP3   & $4.147852\times10^{-6}$ &$1.841478\times10^{10}$ & $7.721912\times10^{10}$ & $1.362103\times10^{11}$ &1.063966 TeV & $0.9957334+i2.751836$ &$7.275089\times10^2$ & $7.76459\times10^{-5}$ \\
				\hline
				BP4   & $1.507167\times10^{-4}$ &$5.200514\times10^{10}$ & $1.630221\times10^{12}$ & $1.753062\times10^{13}$ &10.1 TeV & $-2.247029-i1.65967$ &$4.743545\times10^2$ & $4.028243\times10^{-4}$ \\
				\hline 
		\end{tabular}}
		\caption{Representative benchmark points selected from the viable parameter space obtained in the MCMC scan.}\label{tab:tab0}
	\end{table}
	
	We now proceed to a full exploration of the parameter space using an MCMC scan. The scan ranges used in our analysis are summarized in Table~\ref{tab:tab00}. 
	
	\begin{table}[h]
		\centering
		\begin{tabular}{|c|c|}
			\hline
			Parameter & Scan Range \\
			\hline
			$\log (m_{1}/{\rm eV})$ & -12 to -1 \\
			\hline
			$\log(m_{N_1}/{\rm GeV})$ & 5 to 14 \\
			\hline
			$\log(m_{N_2}/m_{N_1})$ & 0 to 3 \\
			\hline
			$\log(m_{N_3}/m_{N_2})$ & 0 to 3 \\
			\hline
			$\log(m_{\chi}/{\rm GeV})$ & -3 to 10\\
			\hline
			$\log(v_\eta/{\rm GeV})$ & -3 to 2 \\
			\hline
			$a,b$ & -10 to 10 \\
			\hline
			$\log(y_{\chi_{1,2,3}})$ & -6 to 0 \\
			\hline
		\end{tabular}
		\caption{Parameter ranges used in the MCMC scan.}\label{tab:tab00}
	\end{table}
	
	\begin{figure}[h]
		\centering
		\includegraphics[scale=0.38]{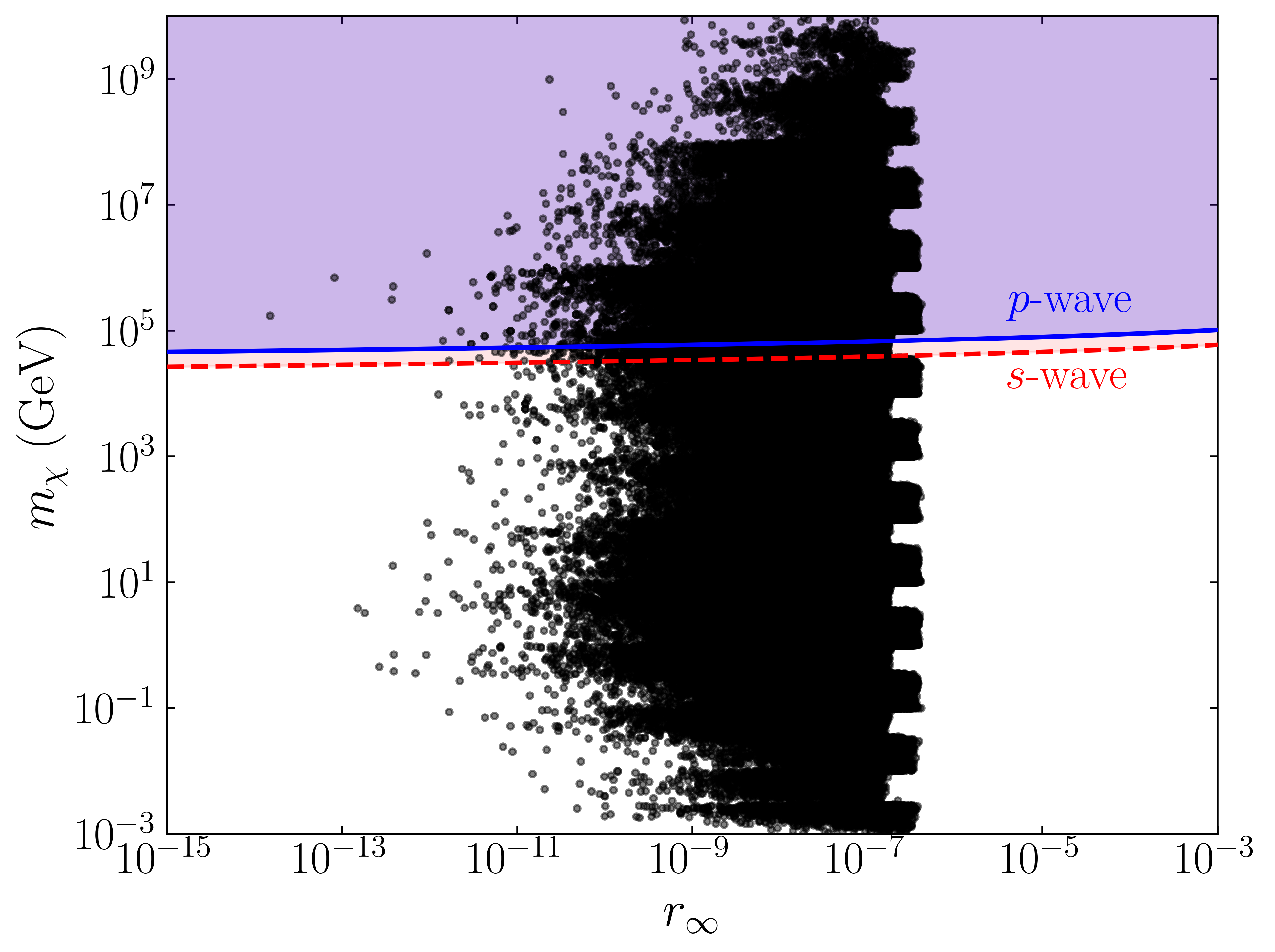}
		\caption{Cogenesis parameter space in the plane of DM mass vs $r_\infty$. The unitarity bounds are shown with red ($s$-wave) and blue ($p$-wave) shaded regions.}
		\label{fig:mxvsrinf}
	\end{figure}
	
	In Fig. \ref{fig:mxvsrinf}, we present the correct lepton asymmetry and correct DM relic points in the plane of $m_\chi$ vs $r_\infty$ with black colored points. We overlay the $s$ and $p$ wave unitary exclusion regions with red and blue colors. We observe that the maximum allowed DM mass from $s$-wave unitarity limit is $\sim39$ TeV, whereas the maximum allowed DM mass from $p$-wave unitarity limit is $\sim68$ TeV. We use 39 TeV as an upper limit on the DM, and in the following discussion, we only keep points with $m_\chi<39$ TeV in the scan plots.
	
	\subsubsection{Full annihilation of the symmetric component of the DM before BBN}
	
	For DM to be fully asymmetric, its symmetric component must efficiently annihilate away. This annihilation can proceed with the help of a singlet scalar $\phi_1$, which also gives rise to sizable DM self-interactions, as will be discussed in the next section. In particular, the symmetric component must annihilate through $\bar{\chi}\chi \rightarrow \phi_1 \phi_1$ before the onset of Big Bang Nucleosynthesis (BBN); otherwise, the late-time production of $\phi_1$ could disrupt the successful predictions of BBN. The annihilation of the symmetric component freezes out when
	\begin{eqnarray}
		n_{\chi}^{\rm eq}(T_{\rm f.o.})\langle\sigma v\rangle\sim \mathcal{H}(T_{\rm f.o.}),
	\end{eqnarray}
	where $T_{\rm f.o.}$ is the freeze out temperature. This leads to 
	\begin{eqnarray}
		\frac{m_\chi}{T_{\rm f.o.}}=\ln\left[ \frac{m_{\rm pl}m_\chi g_\chi\langle\sigma{v}\rangle}{1.66\sqrt{g_*}(2\pi)^{3/2}} \right]+\frac{1}{2}\ln\left[\frac{m_\chi}{T_{\rm f.o.}}\right]
	\end{eqnarray}
	\begin{figure}[h]
		\centering
		\includegraphics[scale=0.38]{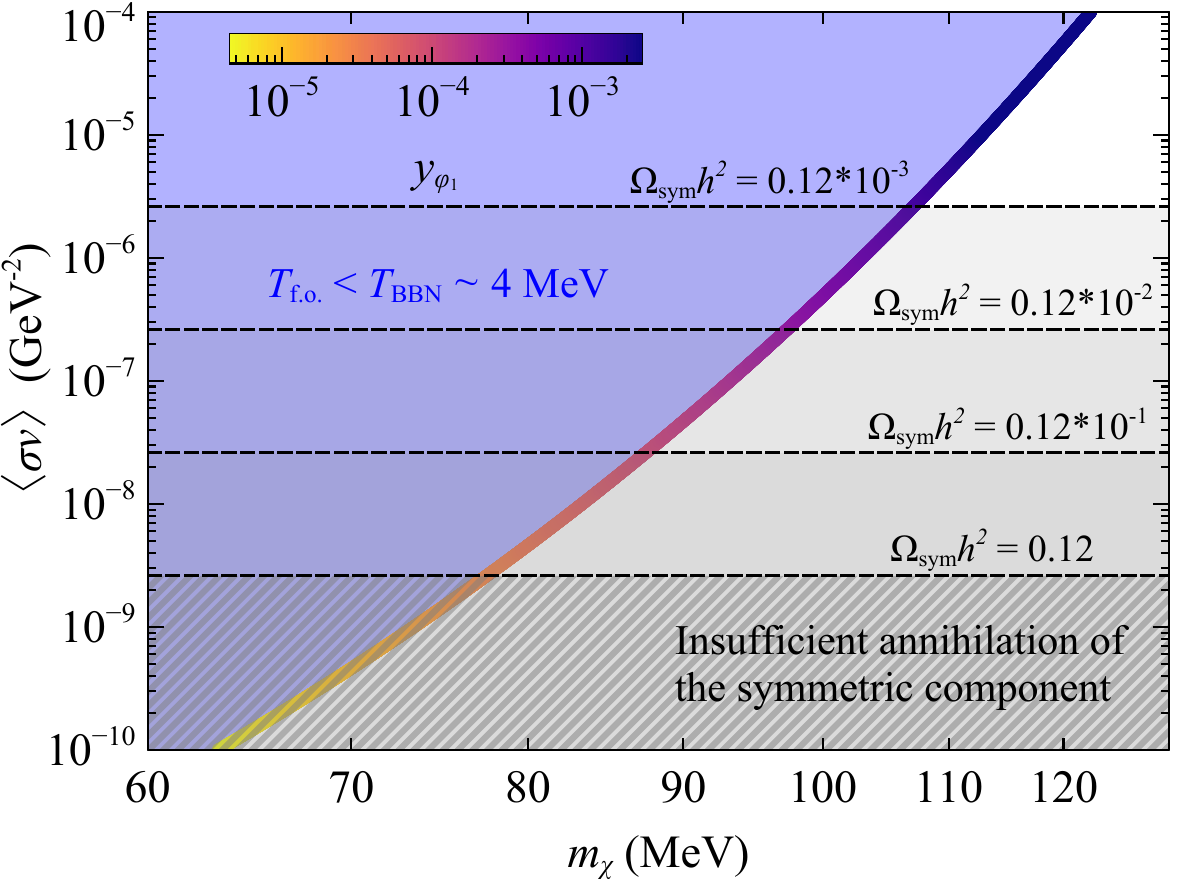}
		\caption{Thermal averaged cross-section as a function of DM mass. In the blue shaded region, the symmetric component of the DM freezes out after the BBN epoch. The annihilation of the symmetric component of the DM is insufficient in the gray shaded region.}
		\label{fig:sigmavVSmdm}
	\end{figure}
	We solve the above equation numerically considering $m_{\rm pl}=1.22\times10^{19}$ GeV, $g_\chi=2$, $g_*\simeq10$ for $T_{\rm f.o.}=T_{\rm BBN}\simeq$ 4 MeV and show the parameter space in the plane of $\langle\sigma v\rangle-m_{\chi}$ in Fig. \ref{fig:sigmavVSmdm}. The blue shaded region is excluded as, in this regime, the symmetric component freezes out after the onset of the BBN epoch. The gray hatched region is also ruled out, since in this region the symmetric component of DM cannot be efficiently annihilated away. The contribution of the symmetric part to the total DM relic can be expressed as
	\begin{eqnarray}
		\Omega_{\rm sym}h^2\simeq\frac{2.6\times10^{-9}~(\rm GeV^{-2})}{\langle\sigma v\rangle}\times0.12.
	\end{eqnarray}
	Different dashed lines represent the amount of symmetric contribution to the total DM relic. The color coding indicates the value of the Yukawa coupling required to obtain the corresponding annihilation cross-section for $\bar{\chi}\chi \rightarrow \phi_1 \phi_1$, arising from the interaction term $y_{\phi_1} \bar{\chi}\chi \phi_1$. We find that DM masses below $\sim 100$ MeV lead to a significant non-vanishing symmetric component. Therefore, in order for DM to be fully asymmetric, its mass must be $\gtrsim 100$ MeV. Consequently, viable ADM can be realized in the mass range $100~\mathrm{MeV} \lesssim m_\chi \lesssim 39~\mathrm{TeV}$. The upper bound arises from unitarity considerations for ADM, as discussed in the previous section. In the following discussion, we consider only DM in the range $100~\mathrm{MeV} \lesssim m_\chi \lesssim 39~\mathrm{TeV}$.
	
	\subsubsection{Results and discussions}
	
	\begin{figure}[h]
		\centering
		\includegraphics[scale=0.37]{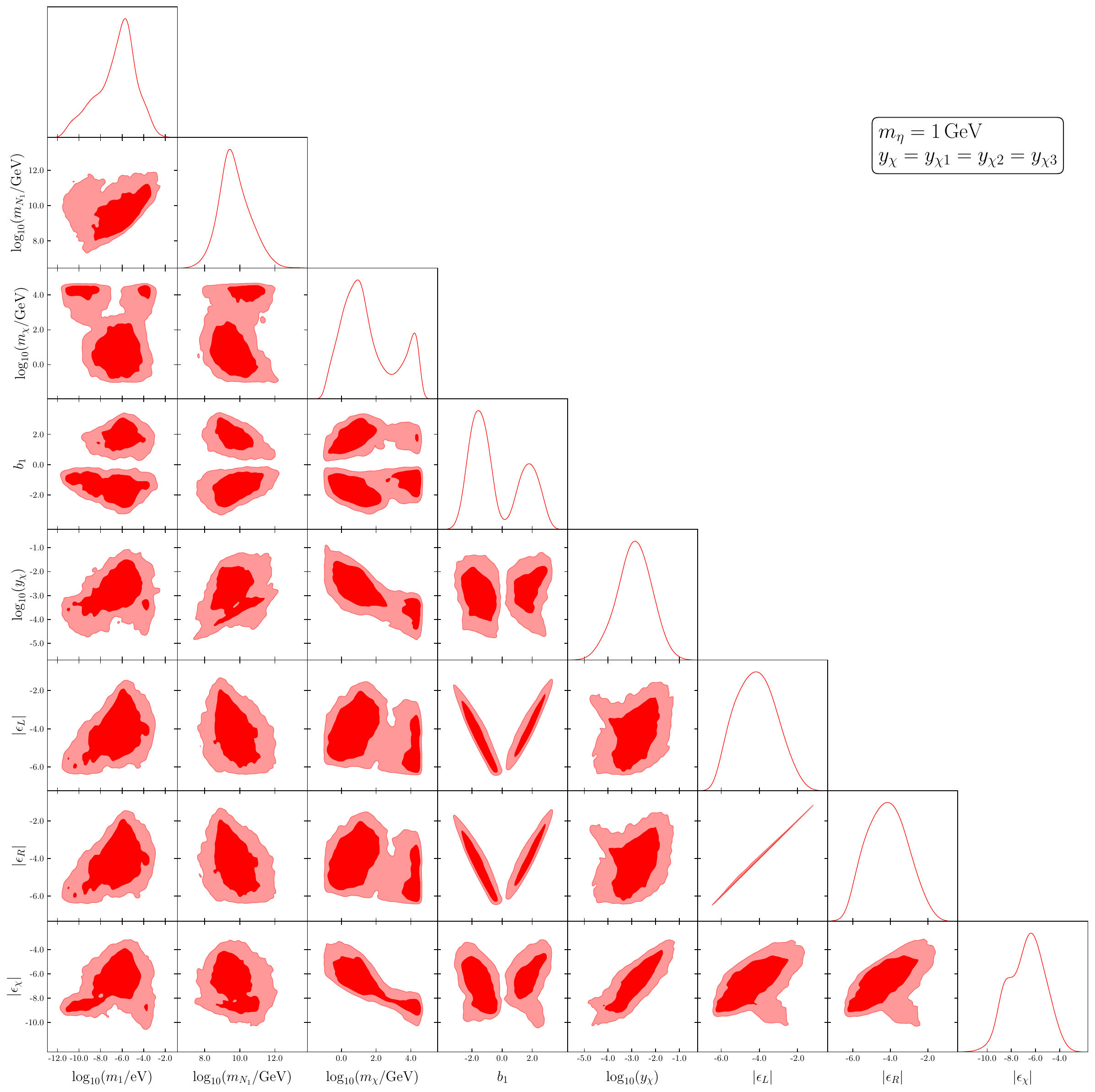}
		\caption{Triangle plot illustrating the correlations among the parameters 
			$m_1$, $m_{N_1}$, $m_\chi$, $b_1$, $y_\chi$ and the resulting $CP$ 
			asymmetries $|\epsilon_L|$, $|\epsilon_R|$, and $|\epsilon_\chi|$ 
			obtained from the MCMC scan. The red shaded regions indicate the 
			$68\%$ (darker) and $95\%$ (lighter) credible regions. We assume
			$m_\eta = 1~\mathrm{GeV}$ and 
			$y_\chi = y_{\chi_1} = y_{\chi_2} = y_{\chi_3}$.}
		\label{fig:trian_CP}
	\end{figure}
	
	In the triangle plot of Fig.~\ref{fig:trian_CP}, we show the correlations among 
	$m_{N_1}$, $m_\chi$, $m_1$, $b$, $y_\chi$, and the corresponding $CP$ asymmetries. 
	We observe that the imaginary part of the Casas--Ibarra rotation angle, $b$, is 
	strongly correlated with the magnitude of the visible sector $CP$ asymmetries, 
	$|\epsilon_L|$ and $|\epsilon_R|$. In particular, the asymmetries increase with 
	increasing values of $b$. This behavior can be understood from the fact that the 
	complex parameter $b$ controls the size of the Yukawa couplings in the 
	Casas--Ibarra parametrization, thereby directly affecting the generated 
	$CP$ asymmetry. We also find a strong positive correlation between the dark sector Yukawa 
	coupling $y_\chi$ and the dark sector $CP$ asymmetry $|\epsilon_\chi|$. 
	Larger values of $y_\chi$ enhance the decay amplitudes contributing to the 
	dark sector asymmetry, leading to the observed positive slope in the 
	$|\epsilon_\chi|-y_\chi$ plane.
	
\begin{figure}[h]
		\centering
		\includegraphics[scale=0.38]{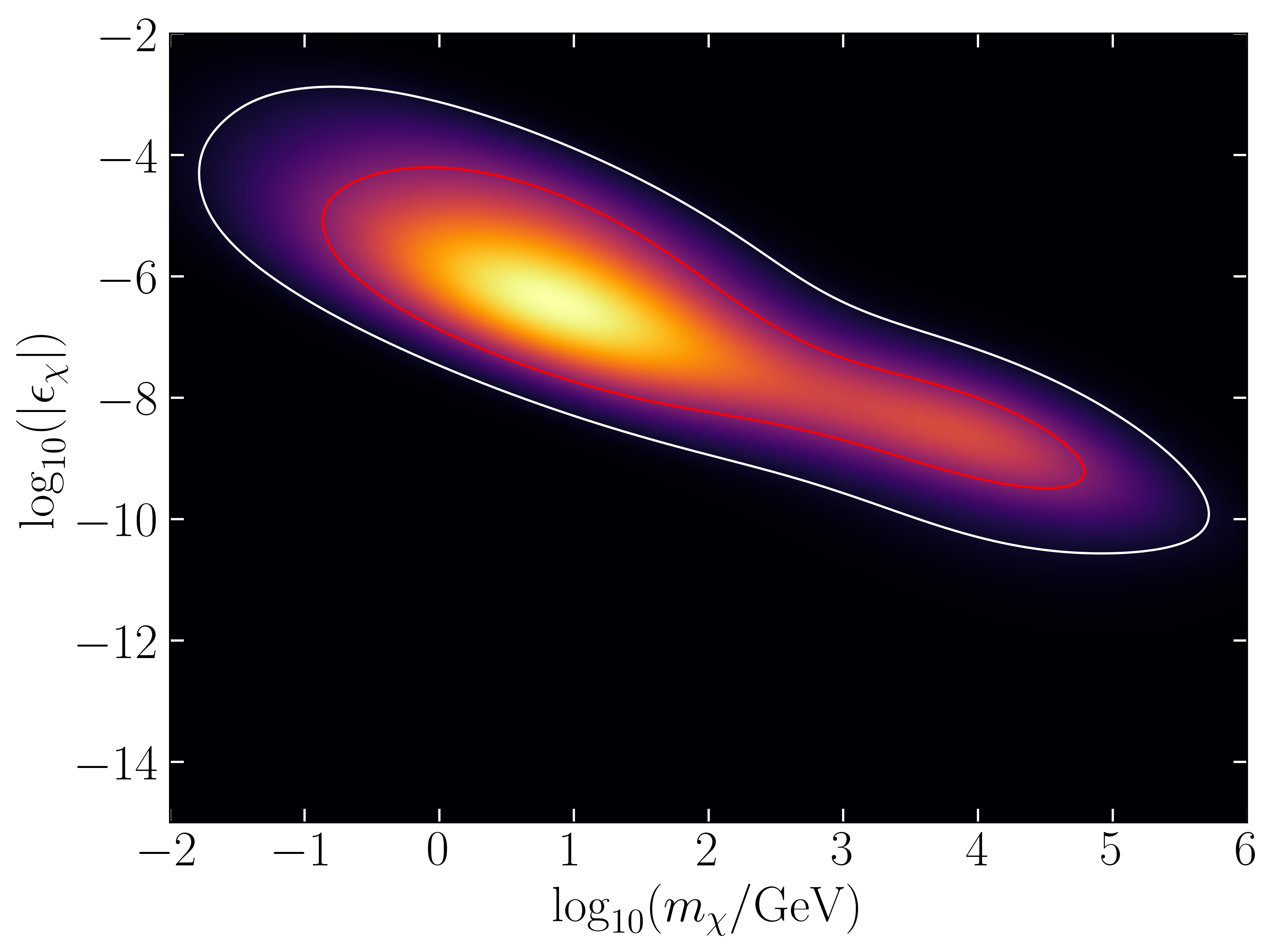}
		\includegraphics[scale=0.38]{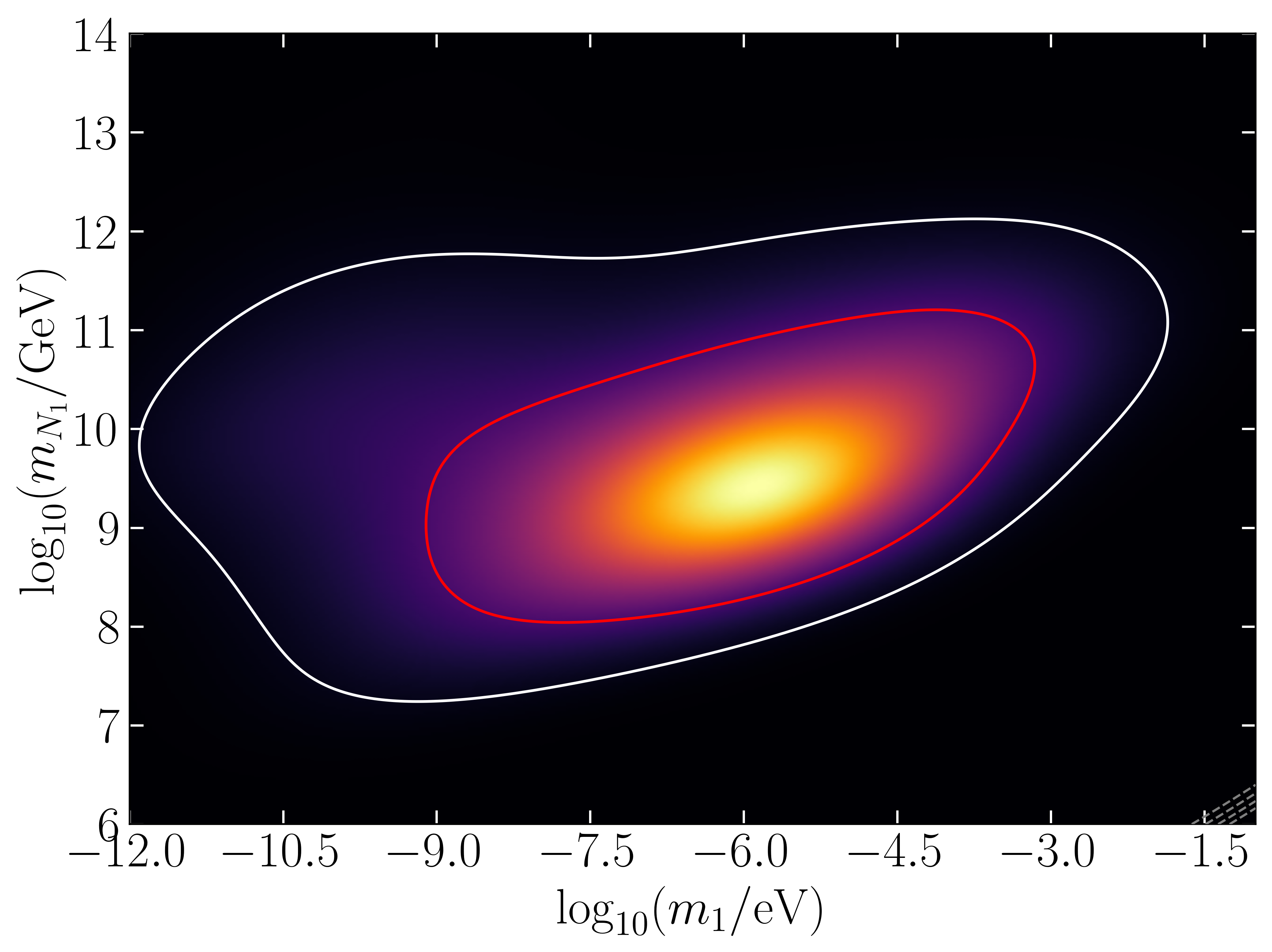}
		\caption{[Left:] 2D credible regions at 68\% (red contour) and 95\% (white contour) C.L. in the $|\epsilon_\chi|-m_\chi$ plane. [Right:] 2D credible regions at 68\% (red contour) and 95\% (white contour) C.L. in the $m_{N_1}-m_1$ plane.}
		\label{fig:scan01}
	\end{figure}

	In the \textit{left} panel of Fig.~\ref{fig:scan01}, we show the $68\%$ (red contour) and $95\%$ (white contour) credible regions in the $|\epsilon_\chi|-m_\chi$ plane. As the DM mass increases, the observed relic abundance can be achieved with a smaller number density of DM particles. Since the DM abundance is directly related to the dark sector \textit{CP} asymmetry $\epsilon_\chi$, a smaller DM number density corresponds to a smaller value of $\epsilon_\chi$. Consequently, $|\epsilon_\chi|$ decreases with increasing DM mass in order to reproduce the correct relic abundance of DM. This trend is clearly visible in the figure.
	
	In the \textit{right} panel of Fig.~\ref{fig:scan01}, we show the $68\%$ (red contour) and $95\%$ (white contour) credible regions in the $m_{N_1}-m_1$ plane. 
	For a fixed set of parameters, increasing $m_1$ leads to a reduction in the generated $CP$ asymmetry while simultaneously enhancing the washout effects. 
	The combined impact of a smaller $CP$ asymmetry and stronger washout can be compensated by increasing the mass of the lightest vector-like fermion, $m_{N_1}$. 
	Since both the visible and dark sector asymmetries originate from the same decay processes, the corresponding $CP$ asymmetries in the two sectors are affected simultaneously. 
	On the other hand, for sufficiently small values of $m_1$, the production rate of the asymmetries becomes suppressed. 
	This suppression can again be compensated by increasing $m_{N_1}$ in order to generate the required asymmetry. 
	This behavior is reflected in the figure, where for $m_1 \gtrsim 10^{-7}~\mathrm{eV}$ the preferred values of $m_{N_1}$ increase.
	
	\begin{figure}[h]
		\centering
		\includegraphics[scale=0.38]{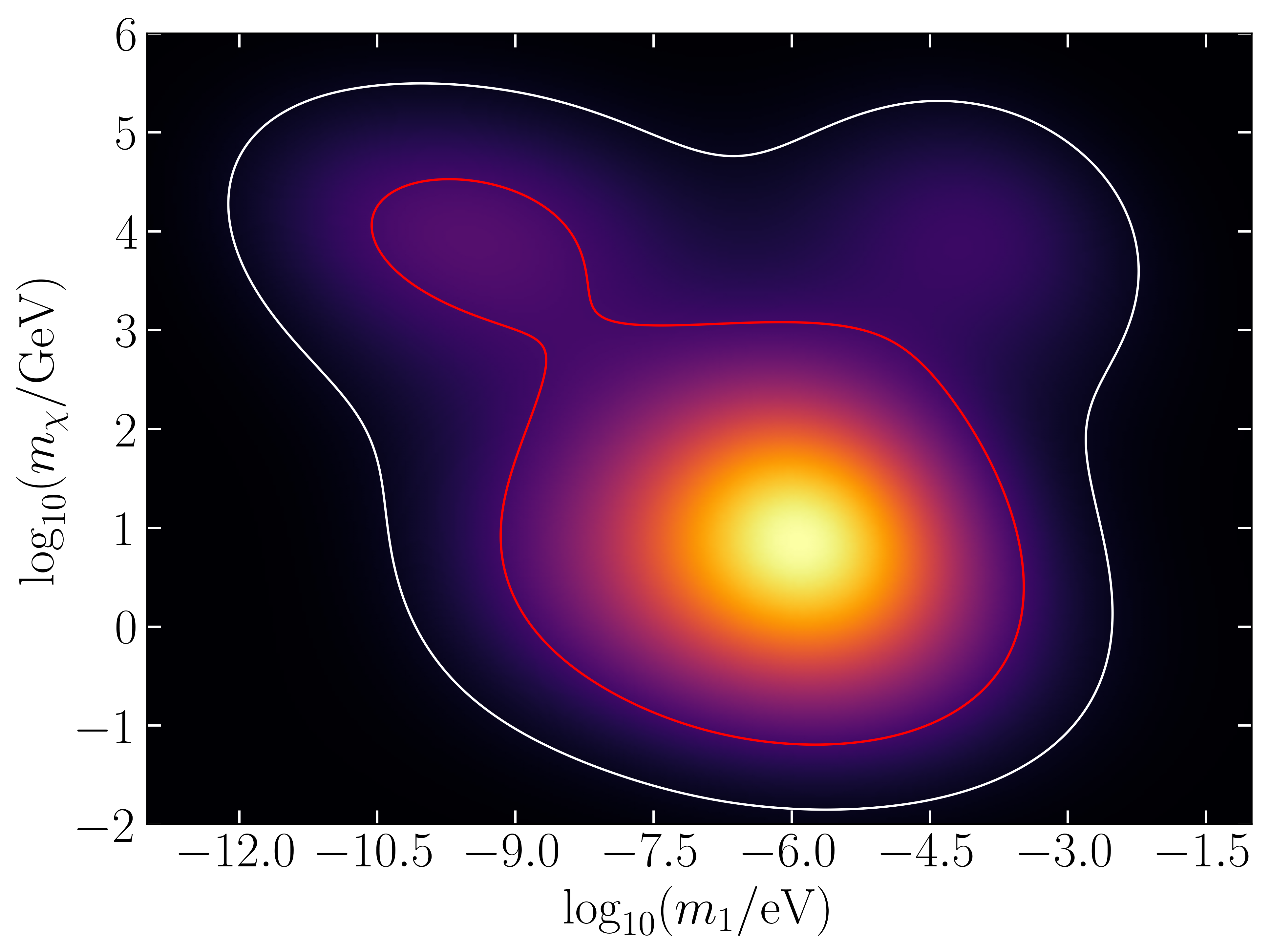}
		\includegraphics[scale=0.38]{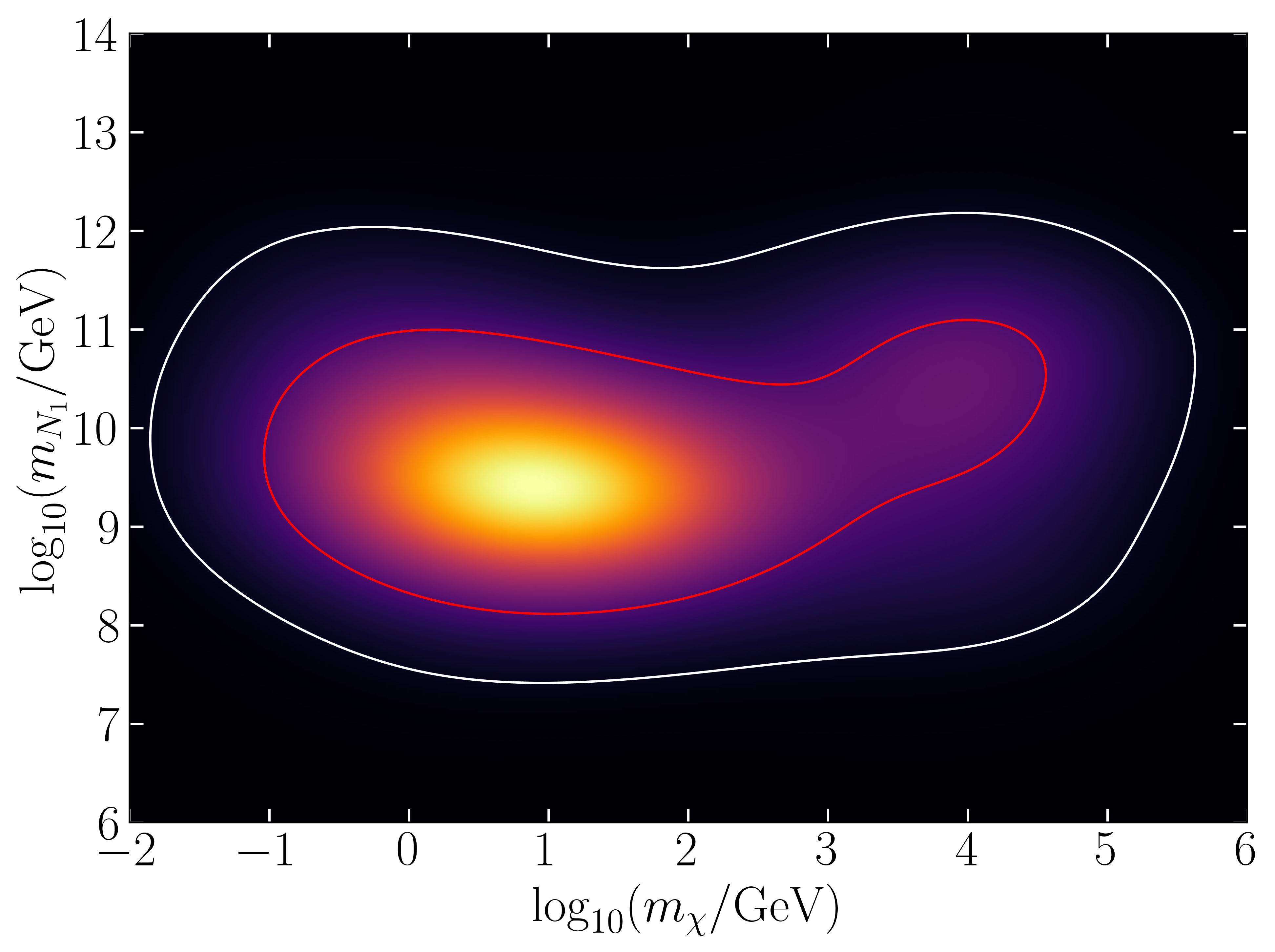}
		\caption{[Left:] 2D credible regions at 68\% (red contour) and 95\% (white contour) C.L. in the $m_{\chi}-m_1$ plane. [Right:] 2D credible regions at 68\% (red contour) and 95\% (white contour) C.L. in the $m_{N_1}-m_\chi$ plane.}
		\label{fig:scan02}
	\end{figure}
	
	In the \textit{left} panel of Fig.~\ref{fig:scan02}, we show the $68\%$ and $95\%$ credible regions (with the same color coding as above) in the $m_\chi$--$m_1$ plane. We find that the preferred range of the lightest neutrino mass is $10^{-10}\,\mathrm{eV} \lesssim m_1 \lesssim 10^{-3}\,\mathrm{eV}$, while the corresponding DM mass spans $10^{-1}\,\mathrm{GeV} \lesssim m_\chi \lesssim 10^{4.6}\,\mathrm{GeV}$. We also observe that smaller values of the lightest neutrino mass, $10^{-11}\,\mathrm{eV} \lesssim m_1 \lesssim 10^{-10}\,\mathrm{eV}$, remain viable provided the DM mass lies in the range $10^{1.5}\,\mathrm{GeV} \lesssim m_\chi \lesssim 10^{4.6}\,\mathrm{GeV}$. In this region of parameter space, assuming normal ordering (NO), the sum of the light neutrino masses is $\sum_{i=1,2,3} m_i \simeq 59~\mathrm{meV}$. This lies well below the current Planck bound \cite{Planck:2018vyg}, but remains within the projected sensitivity of future experiments such as the Simons Observatory, which is expected to probe $\sum_i m_i \lesssim 40~\mathrm{meV}$ at $95\%$ C.L \cite{SimonsObservatory:2019qwx}.
	
	In the \textit{right} panel of Fig.~\ref{fig:scan02}, we present the $68\%$ and $95\%$ credible regions (with the same color coding as above) in the $m_{N_1}$--$m_\chi$ plane. The allowed parameter space spans a wide range of DM masses, $10^{-1}\,\mathrm{GeV} \lesssim m_\chi \lesssim 10^{4.6}\,\mathrm{GeV}$, while the mass of the lightest vector-like fermion is typically found in the interval $10^{7}\,\mathrm{GeV} \lesssim m_{N_1} \lesssim 10^{13}\,\mathrm{GeV}$.

	\section{Detection prospects} 
	\label{sec3}
	
	If the DM particle $\chi$ interacts with a light scalar $\phi_1$, the latter can mediate sizable DM self-interactions and potentially alleviate the small-scale anomalies of the $\Lambda{\rm CDM}$ paradigm \cite{Spergel:1999mh,Buckley:2009in, Feng:2009hw, Feng:2009mn, Loeb:2010gj, Zavala:2012us, Vogelsberger:2012ku}. If such self-interactions are mediated by a light scalar due to their interaction given $-y_{\phi_1}\bar{\chi}\chi\phi_1$, it is also possible to obtain velocity-dependent self-interactions which solve the small-scale issues while being consistent with the collisionless DM picture at large scales \cite{Buckley:2009in, Feng:2009hw, Feng:2009mn, Loeb:2010gj, Bringmann:2016din, Kaplinghat:2015aga, vandenAarssen:2012vpm, Tulin:2013teo}. While such self-interactions can address small-scale structure anomalies, they typically lead to an under-abundant thermal relic for $m_\chi < 10~\mathrm{GeV}$ \cite{Borah:2022ask}. The asymmetric nature of self-interacting DM \cite{Petraki:2014uza, Chen:2023rrl, Dutta:2022knf,Ghosh:2021qbo, Borah:2024wos} can avoid such an under-abundant thermal relic scenario due to chemical potential generated in DM fermion which is prevented from being washed out by keeping it out-of-equilibrium at scales below $T \sim m_{N_1}$ when the asymmetries are generated.
	
	After acquiring the VEV, $\phi_1$ mixes with the SM Higgs, thereby opening up the possibility of spin-independent direct detection (SIDD). The SIDD cross-section is $\sigma^{\rm SI}_{\chi-N}\propto y_{\phi_1}^2\sin^2\gamma/m_{\phi_1}^4$, where $\sin\gamma$ is the $\phi_1-h$ mixing angle. For a typical choice of DM mass, $m_\chi=5$ GeV, $y_{\phi_1}=1$, $\sin\gamma=4\times10^{-7}$, and $m_{\phi_1}=0.25$ GeV, we get the SIDD cross-section to be $\simeq5.75\times10^{-45}~{\rm cm^2}$ which is approximately 1.54 times smaller than the PandaX-4T \cite{PandaX:2025rrz} upper limit on the SIDD cross-section. Cosmological consistency requires the light scalar $\phi_1$ to decay before big bang nucleosynthesis (BBN), i.e., $\tau_{\phi_1} < \tau_{\rm BBN}$, where $\tau_{\phi_1}$ and $\tau_{\rm BBN}$ denote the lifetime of $\phi_1$ and the characteristic BBN timescale, respectively. The lifetime of $\phi_1$ is estimated as
	\begin{eqnarray}
		\tau_{\phi_1}=\left(\sin^2\gamma \frac{m^2_f}{v^2}\frac{m_{\phi_1}}{8\pi}\left( 1-\frac{4m_f^2}{m^2_{\phi_1}} \right)^{3/2}\right)^{-1}.
	\end{eqnarray}
	Now for $m_{\phi_1}=0.25$ GeV, $\sin\gamma=4\times10^{-7}$, and $m_f=m_\mu=105$ MeV, we have $\tau_{\phi_1}=$ 0.0142 s, which is $\ll \tau_{\rm BBN}$. This ensures that the decay of $\phi_1$ does not spoil the successful predictions of primordial nucleosynthesis. This benchmark point satisfies the requirement. The combined requirements from self-interaction strength, direct detection limits, and BBN constraints lead to a nontrivial upper bound on the DM mass, $m_\chi \lesssim 460~\mathrm{GeV}$ \cite{Borah:2024wos}.
	
	Two other prominent detection prospects are the enhanced effective relativistic degrees of freedom $N_{\rm eff}$ \cite{Abazajian:2019oqj} to be probed by cosmic microwave background (CMB) experiments and stochastic gravitational waves (GWs) \cite{Barman:2022yos, Barman:2023fad, King:2023cgv, Borboruah:2024lli,Paul:2024iie, Borah:2026kfo} generated by collapsing domain walls (DWs). The light Dirac neutrinos thermalize with the SM plasma at high temperatures in the early Universe. As the Universe expands and cools down, they decouple from the thermal bath while being relativistic. They contribute to the additional relativistic degrees of freedom: $\Delta{N}_{\rm eff}$ as
	\begin{eqnarray}
		\Delta{N}_{\rm eff}=N_{\nu_R}\left( \frac{g_{*s}(T_{\nu_L}^d)}{g_{*s}(T_{\nu_R}^d)} \right)^{4/3},
	\end{eqnarray}
	\begin{figure}[h]
		\centering
		\includegraphics[scale=0.35]{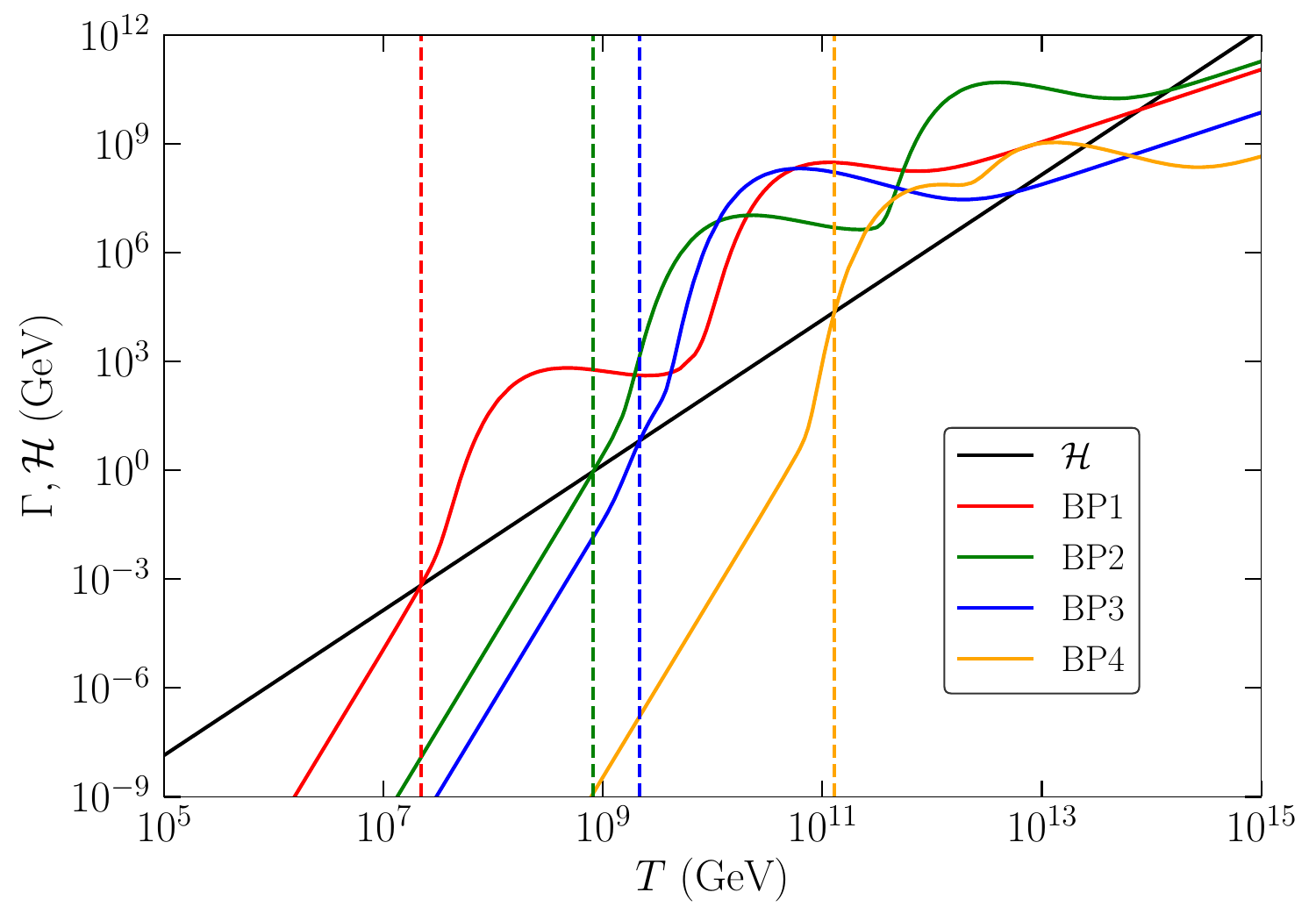}
		\caption{Comparison of Hubble rate with interaction rate keeping $\nu_R$ in equilibrium as a function of temperature for four BPs as given in Table \ref{tab:tab0}. The vertical colored dashed-dotted lines represent the decoupling temperatures for the BPs.}
		\label{fig:GammaT}
	\end{figure} 
	where the $T_{\nu_R}^d$ is the temperature at which $\nu_R$ decouples from the bath and $T_{\nu_L}^d$ is the SM neutrino decoupling temperature which is $\sim$ 1 MeV, $N_{\nu_R}$ = 3 is the number of generations of $\nu_R$. In our case $\nu_R$ decouples for temperatures $T\gg 10^3$ GeV as can be seen in Fig. \ref{fig:GammaT}, where we have shown the comparison of interaction rate for the $s$-channel process $\nu_R\eta\rightarrow\nu_R\eta$ via $N_i$ w.r.t Hubble parameter as a function of $T$ for four BPs as mentioned in Table \ref{tab:tab0}. 
	In this scenario, from Eq 4.2, we get $\Delta{N}_{\rm eff}\simeq3\times(10.6/106.75)^{4/3}\sim 0.14$ \cite{Abazajian:2019oqj} which is within reach of future experiments CMB-S4 \cite{Abazajian:2019eic} and CMB-HD \cite{CMB-HD:2022bsz}. On the other hand, Dirac seesaw models typically involve spontaneously broken discrete symmetries, as outlined in Appendix \ref{app:model}. If the $\eta$ VEV, which is responsible for generating the light Dirac neutrino masses, comes from a spontaneous breaking of a $\mathcal{Z}_2$ symmetry, DWs will form in the early Universe. Their disappearance, with the help of bias terms in the scalar potential, may lead to observable gravitational wave (GW) signatures at present and future GW detectors.

	\section{Conclusion}
	\label{sec4}
	
	We have proposed a novel cogenesis mechanism of dark matter and baryons by extending the minimal type-I seesaw for light Dirac neutrinos by two dark sector particles, namely a Dirac fermion DM $\chi$ and its heavier scalar companion $\phi$. The out-of-equilibrium decay of heavy neutral fermions responsible for the Dirac seesaw leads to the generation of both lepton and dark sector asymmetries. Due to the conservation of total lepton number, the net CP asymmetry vanishes, which can lead to washouts of the asymmetries in the left-handed lepton sector, right-handed neutrino sector, and dark sector if they are brought to equilibrium. Assuming hierarchical heavy neutral fermions, we calculate the CP asymmetries and find the asymmetries in each of these sectors by numerically solving the coupled Boltzmann equations. We show the interesting interplay of asymmetry generation and washouts due to the involvement of three different sectors and constrain the parameter space of the model by performing a statistical analysis and incorporating all theoretical as well as experimental bounds. The symmetric part of DM can annihilate due to a light scalar mediator, which also helps in getting an observable direct-detection rate via the Higgs portal and DM self-interactions. We find that successful cogenesis can be realized for DM masses in the range $100~\mathrm{MeV} \lesssim m_\chi \lesssim 39~\mathrm{TeV}$, while the lightest active neutrino mass remains $\lesssim 10^{-3}$ eV. This amounts to $\sum_{i=1,2,3}m_i\simeq59$ meV. The lower bound on DM mass arises from the requirement that the symmetric component of DM annihilates efficiently before the big bang nucleosynthesis (BBN) epoch, while the upper bound is set by unitarity constraints on the asymmetric DM. The upper bound on the lightest neutrino mass can have implications for experiments sensitive to the effective neutrino mass scale like the KATRIN \cite{KATRIN:2019yun}. Additionally, near-future observation of neutrinoless double beta decay \cite{Dolinski:2019nrj} can also falsify our scenario, as our model predicts a vanishing rate of such a process due to the conservation of global lepton number at the perturbative level. The model also has other interesting detection prospects due to additional dark radiation from light Dirac neutrinos as well as the possibility of the formation of domain walls after spontaneous breaking of discrete symmetries in typical Dirac seesaw models, as illustrated in Appendix \ref{app:model}. Such domain walls can annihilate and emit stochastic gravitational waves, which can be seen in present as well as future experiments. Similarly, the enhanced dark radiation due to light Dirac neutrinos can show up in future CMB measurements. These keep our cogenesis scenario verifiable at experiments operating at different frontiers, including astrophysics, cosmology, and particle physics. 
	
	\acknowledgments The work of D.B. is supported by the Science and Engineering Research Board (SERB), Government of India grant CRG/2022/000603. P.K.P. acknowledges the Ministry of Education, Government of India, for providing financial support for his research via the Prime Minister’s Research Fellowship (PMRF) scheme. P.K.P would like to thank Debarun Paul (ISI, Kolkata) and Biswarup Biswas (University of Turku) for useful discussions. We acknowledge National Supercomputing Mission (NSM) for providing computing resources of `PARAM SEVA' at IIT Hyderabad, which is implemented by C-DAC and supported by the Ministry of Electronics and Information Technology (MeitY) and Department of Science and Technology (DST), Government of India. The authors would like to acknowledge the hospitality by the organizers of WHEPP 2025 workshop at IIT Hyderabad, during which this work was initiated. 
	
	\appendix
	\section{Model realization}\label{app:model}
	Here, we present a model to realize our cogenesis setup. The standard model is extended with three generations of right-handed neutrino $\nu_R$, two generations of vector-like fermion $N_{1,2}$, one vector-like fermion $\chi$, two singlet scalars $\eta$ and $\phi$. We impose a $\mathcal{Z}_4$ symmetry to forbid the Majorana mass of $\nu_R$. We also impose a $\mathcal{Z}_2$ symmetry to forbid the $\bar{L}\tilde{H}\nu_R$ interaction. This $\mathcal{Z}_4$ symmetry is broken softly by the VEV of $\eta$. The DM stability is ensured by $\mathcal{Z}^{\rm dark}_2$ symmetry. The full field content, along with the charge assignments under discrete symmetries, is given in Table \ref{tab:tab1}.
	\begin{table}[h]
		\centering
		\begin{tabular}{|c| c|c| c|} 
			\hline
			Particles & $\mathcal{Z}_4$& $\mathcal{Z}_2$& $\mathcal{Z}^{\rm dark}_2$\\
			\hline
			$L$   & $i$ &+ & + \\
			\hline
			$H$   & $1$ &+ &+ \\
			\hline
			$\nu_R$   & $i$ &-- & + \\
			\hline
			$N$   & $i$ & + &+ \\
			\hline
			$\eta$   & $1$ & -- & + \\
			\hline
			$\chi$   & $i$ & + & -- \\
			\hline
			$\phi$   & $1$ & + & -- \\
			\hline
		\end{tabular}
		\caption{Particles and their charge assignments under $\mathcal{Z}_4\otimes\mathcal{Z}_2\otimes\mathcal{Z}^{\rm dark}_2$ symmetry.}\label{tab:tab1}
	\end{table}
	The relevant Lagrangian is given as
	\begin{eqnarray}
		\mathcal{L}&=&-m_{N_i}\bar{N_i}N_i-m_\chi\bar{\chi}\chi-y_L\bar{L}\tilde{H}N_i- y_R\bar{N_i}\eta\nu_R-y_{\chi_i} \bar{N_i}\chi\phi-V(H,\eta,\phi),
	\end{eqnarray}
	where
	
	\begin{eqnarray}
		V(H,\eta,\phi)&=&-\mu_h^2H^\dagger H+\lambda_h (H^\dagger H)^2+\frac{\mu_\eta^2}{2}\eta^2+\frac{\lambda_\eta}{4}\eta^4+\frac{\mu_\phi^2}{2}\phi^2+\frac{\lambda_\phi}{4}\phi^4
		+\frac{\lambda_{h\eta}}{2}(H^\dagger H)\eta^2\nonumber\\&&+\frac{\mu_1}{\sqrt{2}}(H^\dagger H)\eta +\frac{\lambda_{h\phi}}{2}(H^\dagger H)\phi^2+\frac{\lambda_{\eta\phi}}{4}\eta^2\phi^2 .
	\end{eqnarray}

	\section{Parameterization of Yukawa coupling matrices}\label{app:yukawapara}
	
	The neutrino mass matrix can be written as
	\begin{eqnarray}
		m_\nu=m_L M_N^{-1}m_R,
	\end{eqnarray}
	where $m_L=\frac{y_Lv_h}{\sqrt{2}}$ and $m_R=\frac{y_Rv_\eta}{\sqrt{2}}$.
	
	We can diagonalize the matrix $m_\nu$ by \cite{Casas:2001sr,Cerdeno:2006ha}
	
	\begin{eqnarray}
		\hat{m}_\nu&=&V_L^\dagger m_\nu V_R
	\end{eqnarray}
	\begin{eqnarray}
		\sqrt{\hat{m}_\nu}\sqrt{\hat{m}_\nu}&=&V_L^\dagger m_L M_N^{-1}m_R V_R
	\end{eqnarray}
	\begin{eqnarray}
		\sqrt{\hat{m}_\nu}\sqrt{\hat{m}_\nu}&=&V_L^\dagger m_L \hat{M}_N^{-1}m_R V_R
	\end{eqnarray}
	\begin{eqnarray}
		1&=&\left(\sqrt{\hat{m}^{-1}_\nu}V_L^\dagger m_L \sqrt{\hat{M}_N^{-1}}\right)\left(\sqrt{\hat{M}_N^{-1}}m_R V_R\sqrt{\hat{m}^{-1}_\nu}\right)\nonumber\\
		1&=&A^\dagger B
	\end{eqnarray}
	Now the left-handed mass and Yukawa matrix can be expressed as
	\begin{eqnarray}
		m_L&=&V_L\sqrt{\hat{m}_\nu}A^\dagger\sqrt{\hat{M}_N},\nonumber\\
		y_L&=&\frac{\sqrt{2}}{v_h}V_L\sqrt{\hat{m}_\nu}A^\dagger\sqrt{\hat{M}_N},
	\end{eqnarray}
	and the right-handed mass and Yukawa matrix can be expressed as
	\begin{eqnarray}
		m_R&=&\sqrt{\hat{M}_N}B\sqrt{\hat{m}_\nu}V_R^\dagger,\nonumber\\
		y_R&=&\frac{\sqrt{2}}{v_\eta}\sqrt{\hat{M}_N}B\sqrt{\hat{m}_\nu}V_R^\dagger.
	\end{eqnarray}
	Here $\hat{m}_\nu={\rm diag}(m_1,m_2,m_3)$ and $\hat{M}_N={\rm diag}(m_{N_1},m_{N_2},m_{N_3})$. We use the values of neutrino oscillation parameters at their best fit values \cite{deSalas:2020pgw}, considering normal ordering in our analysis. For simplicity, we choose $V_L=U_{\rm PMNS}$,  $V_R=I_{3\times3}$, and $A=B=R$ and a general rotation matrix with complex rotation angle $\theta=a+ib$.

	\section{Relevant cross-sections and decay widths}\label{app:crossec}
	The decay width of $N_i$ to the visible left-handed sector is given as
	\begin{eqnarray}
		\Gamma_{N_i\rightarrow LH}=\frac{(y_L^\dagger y_L)_{ii}}{32\pi}m_{N_i}
	\end{eqnarray}
	The decay width of $N_i$ to the visible right-handed sector is given as
	\begin{eqnarray}
		\Gamma_{N_i\rightarrow \eta\nu_R}=\frac{(y_R^\dagger y_R)_{ii}}{64\pi}m_{N_i}\left( 1-\frac{m_\eta^2}{m_{N_i}^2} \right)^2
	\end{eqnarray}
	The decay width of $N_i$ to the dark sector is given by
	\begin{eqnarray}
		\Gamma_{N_i\rightarrow \chi\phi}&=&\frac{(y_{\chi i}y^*_{\chi i})}{16\pi}m_{N_i}\left( (1+x)^2-y^2 \right) \left( x^4+(1-y^2)^2-2x^2(1+y^2) \right)^{1/2},
	\end{eqnarray}
	where $x=m_{\chi}/m_{N_i}$ and $y=m_{\phi}/m_{N_i}$.
	
	The $2\leftrightarrow2$ scattering cross-sections entering into the Boltzmann equations are given as
	\begin{eqnarray}  \sigma(\chi\phi\rightarrow\nu_R\eta)&=&\frac{(y_R^\dagger y_R)_{ii}y_{\chi i}y^*_{\chi i}}{128\pi}\frac{E^2}{s^2A^2} \frac{(4sm_\chi m_{N_i}+m_\chi^2 B+CB)}{\sqrt{m_\chi^4+C^2-2m_\chi^2D}},
	\end{eqnarray}
	where $A=s-m_{N_i}^2,B=s+m_{N_i}^2,C=s-m_{\phi}^2,D=s+m_{\phi}^2,E=s-m_{\eta}^2$.
	
	\begin{eqnarray}
		\sigma(\nu_R\eta\rightarrow LH)=\frac{(y_R^\dagger y_R)_{ii}(y_L^\dagger y_L)_{ii}}{128\pi}\frac{m_{N_i}^2}{(s-m_{N_i}^2)^2}
	\end{eqnarray}
	
	\begin{eqnarray}
		\sigma(\chi\phi\rightarrow LH)&=&\frac{(y_L^\dagger y_L)_{ii}y_{\chi i}y^*_{\chi i}}{64\pi}\frac{1}{A^2}
		\frac{(4sm_\chi m_{N_i}+m_\chi^2 B+CB)}{\sqrt{m_\chi^4+C^2-2m_\chi^2D}}
	\end{eqnarray}
	
	The thermally averaged reaction density can be computed for any process $ab\rightarrow cd$ as
	\begin{eqnarray}
		\gamma_{ab\rightarrow cd}&=&\frac{T}{32\pi^4}g_ag_b\int_{(m_a+m_b)^2}^\infty \sigma_{ab\rightarrow cd}(s)\sqrt{s}K_1\left(\frac{\sqrt{s}}{T}\right)\nonumber\\&&
		\frac{\left( s-(m_a+m_b)^2 \right)\left( s-(m_a-m_b)^2 \right)}{s}ds
	\end{eqnarray}
	and for any decay process $a\rightarrow bc$, it is defined as
	\begin{eqnarray}
		\gamma_{a\rightarrow bc}=n_a^{\rm eq}\frac{K_1(m_a/T)}{K_2(m_a/T)}\Gamma_{a\rightarrow bc}.
	\end{eqnarray}
	Here, $K_i$s are the modified Bessel functions, $g_i$s are the degrees of freedom of any species, and $n_i^{\rm eq}$s are the equilibrium number densities.
	
	\section{Upper bound on $m_1$}\label{app:m1up}
	In our analysis we have assumed $V_L=U_{\rm PMNS}$, $V_R=I_{3\times3}$ and $A=B=R$. With these simplifications we have
	\begin{eqnarray}
		y_L=\frac{\sqrt{2}}{v_h}U_{\rm PMNS}\sqrt{\hat{m}_\nu}R^\dagger\sqrt{\hat{M}_N},~~y_R=\frac{\sqrt{2}}{v_\eta}\sqrt{\hat{M}_N}R\sqrt{\hat{m}_\nu}.
	\end{eqnarray}
	Now
	\begin{eqnarray}
		(y_L^\dagger y_L)_{11}&=&\frac{2}{v_h^2}\left( \sqrt{\hat{M}_N} R \sqrt{\hat{m}_\nu} U_{\rm PMNS}^\dagger U_{\rm PMNS} \sqrt{\hat{m}_\nu} R^\dagger \sqrt{\hat{M}_N} \right)_{11}\nonumber\\&=&\frac{2}{v_h^2}\left( \sqrt{\hat{M}_N} R {\hat{m}_\nu} R^\dagger \sqrt{\hat{M}_N} \right)_{11}=\frac{2}{v_h^2}\left( \sqrt{\hat{M}_N}\mathcal{M}\sqrt{\hat{M}_N} \right)_{11}\nonumber\\&=&\frac{2}{v_h^2} m_{N_1}\mathcal{M}_{11},
	\end{eqnarray}
	where
	\begin{eqnarray}
		\mathcal{M}_{11}&=&\left( R\hat{m}_\nu R^\dagger \right)_{11}=\sum_k\sum_l R_{1k}{(\hat{m}_{\nu})}_{kl}R^\dagger_{l1}=\sum_k\sum_l R_{1k}m_k\delta_{kl}R^\dagger_{l1}\nonumber\\
		&=&\sum_kR_{1k}m_kR^\dagger_{k1}=\sum_kR_{1k}m_kR^*_{1k}=\sum_km_k|R_{1k}|^2.
	\end{eqnarray}
	Therefore,
	\begin{eqnarray}
		(y_L^\dagger y_L)_{11}&=&\frac{2}{v_h^2} m_{N_1}\sum_km_k|R_{1k}|^2
	\end{eqnarray}
	Similarly 
	\begin{eqnarray}
		(y_R^\dagger y_R)_{11}&=&\frac{2}{v_\eta^2}\left( \sqrt{\hat{m}_\nu} R^\dagger \sqrt{\hat{M}_N}\sqrt{\hat{M}_N} R \sqrt{\hat{m}_\nu}\right)_{11}\nonumber\\&=&\frac{2}{v_\eta^2}\left( \sqrt{\hat{m}_\nu} R^\dagger{\hat{M}_N} R \sqrt{\hat{m}_\nu}\right)_{11}=\frac{2}{v_\eta^2} m_{1}\mathcal{M}^\prime_{11},\label{eq:yryr11}
	\end{eqnarray}
	where
	\begin{eqnarray}
		\mathcal{M}^\prime_{11}&=&\left( R^\dagger\hat{M}_N R \right)_{11}=\sum_k m_{N_k}|R_{k1}|^2.
	\end{eqnarray}
	\begin{eqnarray}
		(y_R^\dagger y_R)_{11}&=&\frac{2}{v_\eta^2} m_{1}\sum_km_{N_k}|R_{k1}|^2
	\end{eqnarray}
	The decay parameter for the left-handed sector can be estimated to be
	\begin{eqnarray} K_L&=&\frac{\Gamma_{N_1\rightarrow LH}}{\mathcal{H}}\simeq \frac{(y_L^\dagger y_L)_{11}m_{N_1}/32\pi}{1.66\sqrt{g_*}m_{N_1}^2/m_{\rm pl}};~~~{\rm taking } ~T\sim m_{N_1}\nonumber\\
		&=&\frac{\mathcal{M}_{11} m_{\rm pl}}{16\pi 1.66\sqrt{g_*} v_h^2}\simeq23 ~~{\rm with}~~\mathcal{M}_{11}=0.1~{\rm eV},~g_*=106.75.
	\end{eqnarray}
	The left and right-handed sectors should not thermalize. This is ensured by the requirement
	\begin{eqnarray}
		\frac{\Gamma_{\nu_R\eta\leftrightarrow LH}}{\mathcal{H}}<1,\label{eq:LReq}
	\end{eqnarray}
	where $\Gamma_{\nu_R\eta\leftrightarrow LH}$ can be estimated as
	\begin{eqnarray} \Gamma_{\nu_R\eta\leftrightarrow LH}&=&n_{\rm eq}\langle\sigma{v}\rangle\sim T^3\frac{(y_L^\dagger y_L)_{11}(y_R^\dagger y_R)_{11}}{128\pi m_{N_1}^2}=m_{N_1}^3\frac{(y_L^\dagger y_L)_{11}(y_R^\dagger y_R)_{11}}{128\pi m_{N_1}^2}\nonumber\\&=&\frac{(y_L^\dagger y_L)_{11}(y_R^\dagger y_R)_{11}}{128\pi}m_{N_1}.
	\end{eqnarray}
	From Eq. \eqref{eq:LReq},
	\begin{eqnarray}
		\frac{\Gamma_{\nu_R\eta\leftrightarrow LH}}{\mathcal{H}}&<&1\nonumber\\
		\frac{\Gamma_{\nu_R\eta\leftrightarrow LH}}{\Gamma_{N_1\rightarrow LH}}\frac{\Gamma_{N_1\rightarrow LH}}{\mathcal{H}}&<&1\nonumber\\
		\frac{(y_R^\dagger y_R)_{11}}{4}K_L&<&1
	\end{eqnarray}
	Now using Eq. \eqref{eq:yryr11} in the above expression, we get
	\begin{eqnarray}
		m_1<\frac{2v_\eta^2}{\mathcal{M}^\prime_{11} K_L}
	\end{eqnarray}
	\begin{eqnarray}
		m_1\lesssim 2\times10^{-3}~{\rm eV} \left( \frac{v_\eta}{100 {~\rm GeV}} \right)^2\left( \frac{10^{15} {~\rm GeV}}{\mathcal{M}^\prime_{11}} \right) \left( \frac{10}{K_L} \right).
	\end{eqnarray}

	\section{MCMC analysis}\label{app:mcmc}
	
	We explore the multidimensional parameter space of the model using a Markov Chain Monte Carlo (MCMC) technique, following the standard Bayesian framework \cite{Hinkley:1969fzb,Arina:2011cu}. The MCMC scan is performed to identify regions of parameter space consistent with the observed baryon asymmetry of the Universe and the measured dark matter relic abundance. The likelihood associated with the baryon asymmetry is taken to be Gaussian and is defined as
	\begin{eqnarray}
		\ln\mathcal{L}_{\eta_B}=-\frac{1}{2}\left( \frac{\eta_B-\overline{\eta_B}}{\sigma_{\eta_B}} \right)^2
	\end{eqnarray}
	where $\eta_B$ denotes the baryon-to-photon ratio predicted by the model. We use $\overline{\eta_B}\pm\sigma_{\eta_B}=(6.0\pm0.5)\times10^{-10}$.
	
	To incorporate the constraint on the DM relic abundance while allowing for a non-trivial relation between the DM and baryon abundances, we employ the likelihood function introduced in Ref.~\cite{Arina:2011cu}. This likelihood effectively marginalizes over the ratio $r \equiv \Omega_{\rm DM}/\Omega_b$ and is given by
	\begin{eqnarray}
		\mathcal{L}_{\Omega_{\rm DM} h^2}&=&\frac{1}{\sqrt{2\pi}\sigma_b\sigma_{\rm DM}}\frac{b(r)c(r)}{a(r)^3}\Phi\left(\frac{b(r)}{a(r)}\right)\frac{{\rm exp}\left(-\frac{1}{2}\left(\overline{(\Omega_{\rm DM}h^2)^2}/\sigma_{\rm DM}^2+\overline{(\Omega_{b}h^2)^2}/\sigma_{b}^2\right)\right)}{a(r)^2\pi\sigma_b\sigma_{\rm DM}},\nonumber\\
	\end{eqnarray}
	where the functions $a(r)$, $b(r)$, and $c(r)$ are defined as
	\begin{eqnarray}
		a(r)&=&\sqrt{\frac{r^2}{\sigma_{\rm DM}^2}+\frac{1}{\sigma_b^2}},\nonumber\\
		b(r)&=&\frac{\overline{\Omega_{\rm DM}h^2}}{\sigma_{\rm DM}^2}r+\frac{\overline{\Omega_{b}h^2}}{\sigma_{b}^2},\nonumber\\
		c(r)&=&{\rm exp}\left( \frac{1}{2}\frac{b(r)^2}{a(r)^2}-\frac{1}{2} \left(\frac{\overline{(\Omega_{\rm DM}h^2)^2}}{\sigma_{\rm DM}^2}+\frac{\overline{(\Omega_{b}h^2)^2}}{\sigma_{b}^2}  \right)  \right),\nonumber\\
		\Phi(u)&=&{\rm Erf}\left(\frac{u}{\sqrt{2}}\right).
	\end{eqnarray}
	We adopt the Planck 2018 \cite{Planck:2018vyg} values for the present-day energy densities,
	$\overline{\Omega_{\rm DM} h^2}\pm\sigma_{\rm DM}=0.1200 \pm 0.0012 $. $\overline{\Omega_{b} h^2}\pm\sigma_{b}=0.02237 \pm 0.00015 $. Assuming statistical independence between the baryon asymmetry and DM abundance measurements, the total log-likelihood used in the MCMC analysis is
	\begin{eqnarray}
		\ln\mathcal{L}_{\rm cogenesis}=\ln\mathcal{L}_{\eta_B}+\ln\mathcal{L}_{\Omega_{\rm DM} h^2}.
	\end{eqnarray}
	
	This combined likelihood is used to assess the viability of parameter points and to extract the preferred regions consistent with successful cogenesis.
	\begin{figure}[H]
		\centering    \includegraphics[scale=0.35]{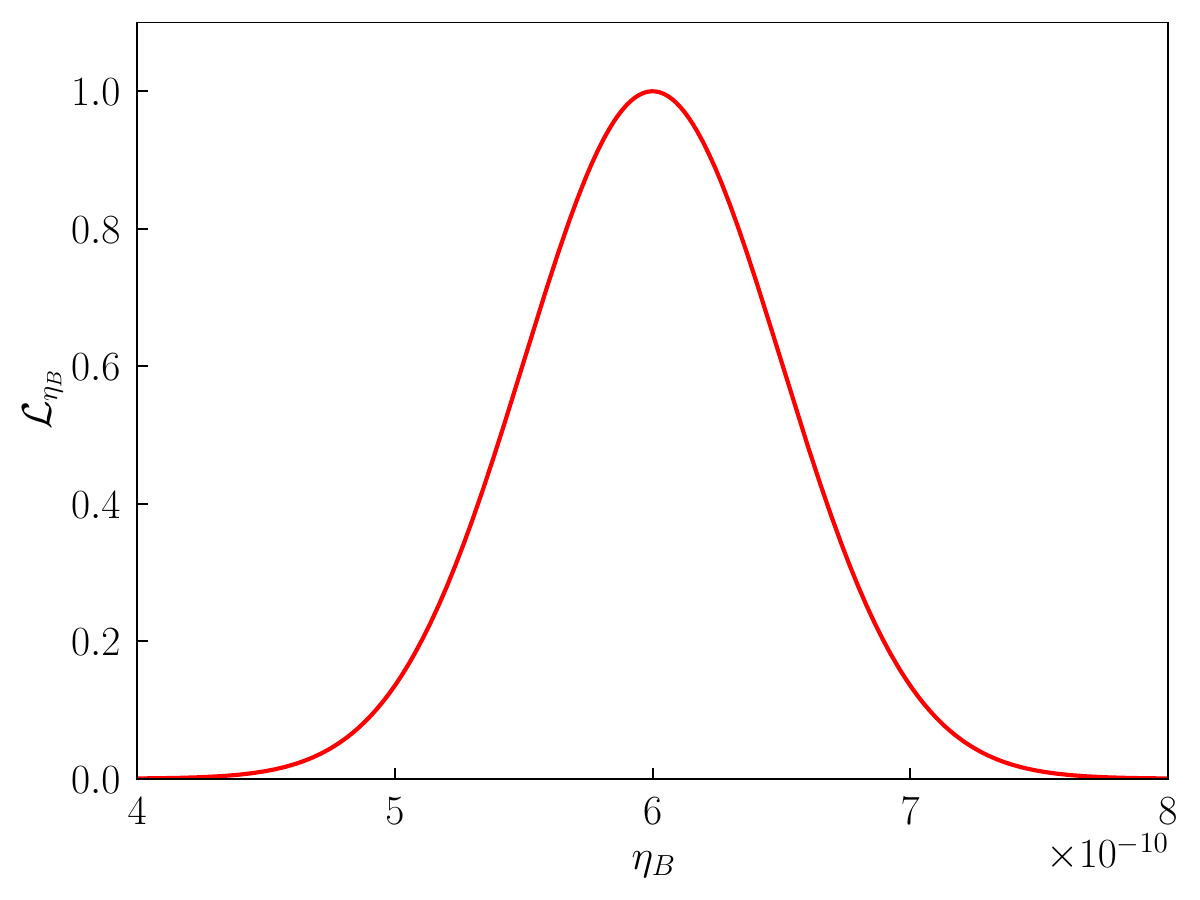}
		\includegraphics[scale=0.35]{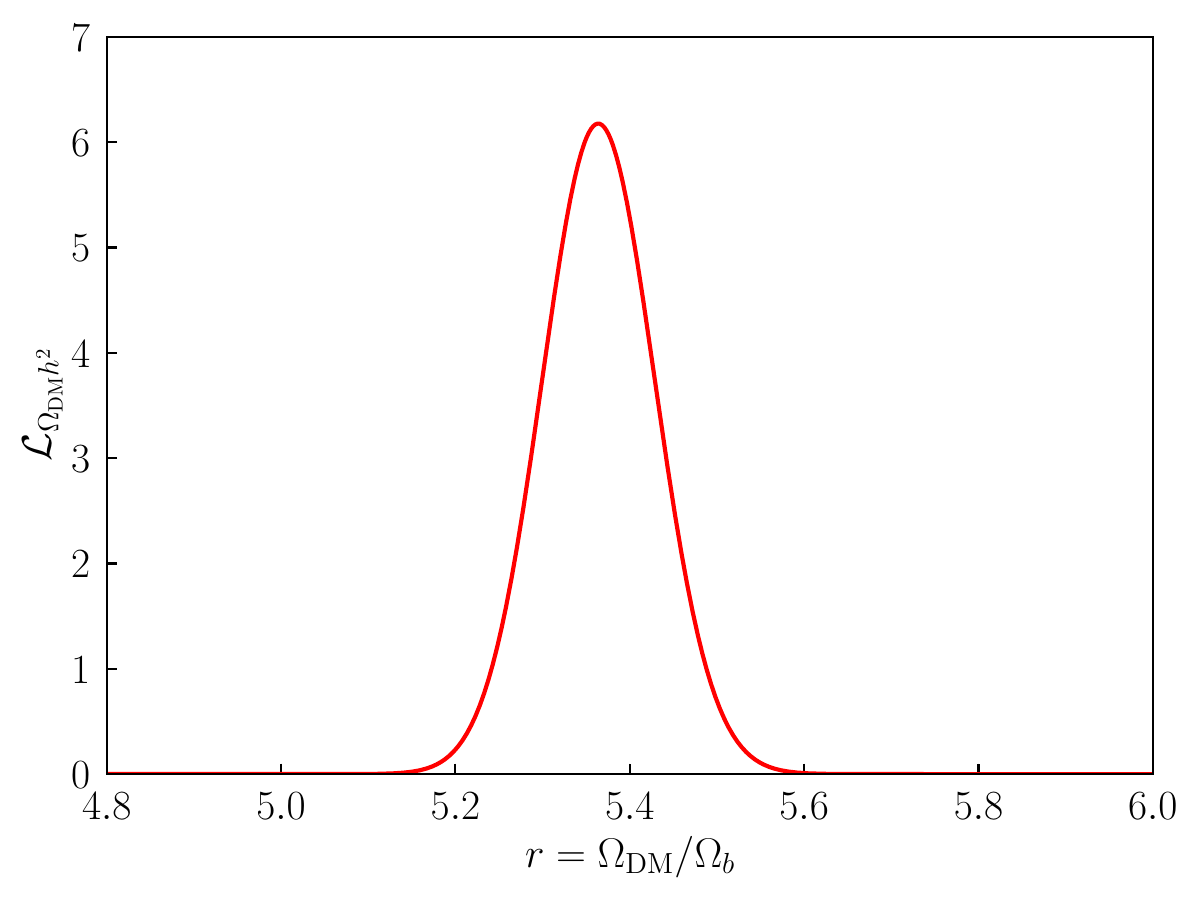}
		\caption{[\textit{Left}]: Likelihood function $\mathcal{L}_{\eta_B}$ as a function of $\eta_B$. [\textit{Right}]: Likelihood function $\mathcal{L}_{\Omega_{\rm DM}h^2}$ as a function of $r=\Omega_{\rm DM}/\Omega_b$.}
		\label{fig:likelihood}
	\end{figure}
	In the \textit{left} panel of Fig.~\ref{fig:likelihood}, we show the likelihood function $\mathcal{L}_{\eta_B}$ as a function of $\eta_B$. The \textit{right} panel displays the likelihood function $\mathcal{L}_{\Omega_{\rm DM}h^2}$ as a function of $r=\Omega_{\rm DM}/\Omega_b$. In both cases, the likelihood peaks around the expected values, $\eta_B \simeq 6\times10^{-10}$ and $r \simeq 5.36$.
	\begin{figure}[tbh]
		\centering    \includegraphics[scale=0.27]{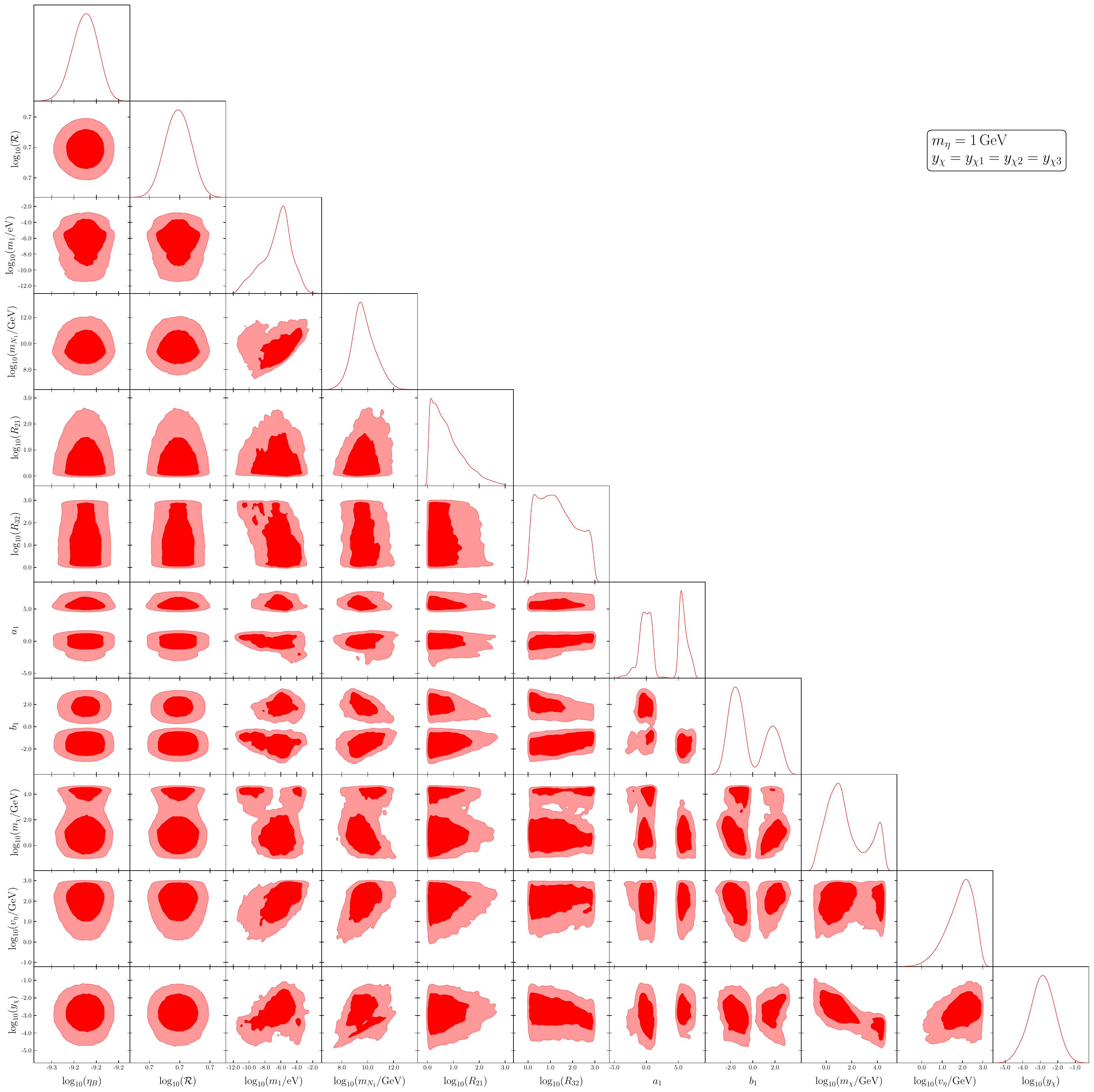}
		\caption{Triangle plot showing the correlations among the relevant model parameters obtained from the MCMC scan. The darker (lighter) regions correspond to the 68\% (95\%) credible regions.}
		\label{fig:triangle_appendix}
	\end{figure}
	For completeness, we present in Fig.~\ref{fig:triangle_appendix} the full correlation matrix among the model parameters obtained from the MCMC scan.


\providecommand{\href}[2]{#2}\begingroup\raggedright\endgroup

\end{document}